\documentclass{degruyter-proceedings}          
\newcommand{\vek}[1]{\bf{#1}}
\newcommand{\bea}{\begin{eqnarray}} 
\newcommand{\eea}{\end{eqnarray}} 
\newcommand{\nn}{\nonumber} 

\title{Observations of General Relativity at strong and weak limits}
\headlinetitle{Strong and Weak GR}

\lastnameone{Byrd}
\firstnameone{Gene G.}
\nameshortone{G.\,G.~Byrd}
\addressone{University of Alabama, Tuscaloosa, AL, 35487-0324}
\countryone{USA}
\emailone{genebyrd@bellsouth.net}

\lastnametwo{Chernin}
\firstnametwo{Arthur}
\nameshorttwo{A.\,D.~Chernin}
\addresstwo{Sternberg Astronomical Institute, Moscow University, 119899, Moscow}
\countrytwo{Russia}
\emailtwo{arthur.chernin@gmail.com}

\lastnamethree{Teerikorpi}
\firstnamethree{Pekka}
\nameshortthree{P.\,~Teerikorpi}
\addressthree{Tuorla Observatory, Department of Physics and Astronomy, University of Turku, 21500, Piikki\"o}
\countrythree{Finland}
\emailthree{pekkatee@utu.fi}

\lastnamefour{Valtonen}
\firstnamefour{Mauri}\nameshortfour{M.\,J.~Valtonen}
\addressfour{FINCA, University of Turku, FI-21500 Piikki\"o}
\countryfour{Finland}
\emailfour{mvaltonen2001@yahoo.com}

\lastnamefive{}
\firstnamefive{}
\nameshortfive{}
\addressfive{}
\countryfive{}
\emailfive{}

\abstract{Einstein's General Relativity theory has been tested in many ways during the last hundred years as reviewed in this chapter. Two tests are discussed in detail in this article: the concept of a zero gravity surface, the roots of which go back to J\"arnefelt, Einstein and Straus, and the no-hair theorem of black holes, first proposed by Israel, Carter and Hawking. The former tests the necessity of the cosmological constant $\Lambda$, the latter the concept of a spinning black hole. The zero gravity surface is manifested most prominently in the motions of dwarf galaxies around the Local Group of galaxies. The no-hair theorem is testable for the first time in the binary black hole system OJ287. These represent stringent tests at the limit of weak and strong gravitational fields, respectively. In this article we discuss the current observational situation and future possibilities.}



\researchsupported{This work has been supported by The Finnish Society of Sciences and Letters and The Finnish Academy of Science and Letters.}

\acknowledgments{We note that Bruce Barker (who died in 1988 at the age of 54) was a co-author of two important theoretical papers testing General Relativity in extreme binary systems like those discussed here.  He was a friend and colleague of M. V. and G. B. at the University of Alabama.}

\firstpage{1}

\begin{document}


\section{Introduction}

In his theory of General Relativity, Einstein (1916) concluded that matter causes curvature in the surrounding spacetime, and bodies react to this curvature in such a way that there appears to be a gravitational attraction which causes acceleration.  From the geometry of spacetime, it is possible to calculate the orbits of bodies which are influenced by gravity.  In flat spacetime the force-free motion happens on a straight line, but in a spacetime curved by mass/energy the force free motion can create practically closed orbits as seen in the elliptical motion of a planet around the Sun.

 The Einstein equations with the cosmological constant $\Lambda$ (Einstein 1917) have the form: 
\begin{eqnarray}
R_{\mu\nu}-\frac{1}{2}Rg_{\mu\nu}+\Lambda g_{\mu\nu}=-\frac{8\pi G}{c^4}T_{\mu\nu}.
\end{eqnarray}
The Ricci tensor $R_{\mu\nu}$ and the Ricci scalar $R$ are functions of the metric tensor $g_{\mu\nu}$. The metric tensor describes the geometry of the spacetime while the Ricci tensor and the Ricci scalar measure its curvature. The energy momentum tensor $T_{\mu\nu}$ has the dominant component $T_{00}=\rho_{M}c^{2}$ where $\rho_{M}$ is the matter density, $G$ is the gravitational constant and $c$ is the speed of light. See e.g. Byrd et al. (2007, 2012) for details.

In the presence of the cosmological constant $\Lambda$ the metric of a spherically symmetric object of mass $M$ is (Lema\^itre 1931, McVittie 1932, 1933)

\begin{eqnarray}
ds^2=-\biggl(1-\frac{2 GM}{rc^2}-\frac{\Lambda r^2}{3}\biggr)dt^2
+\biggl(1-\frac{2 GM}{rc^2}-\frac{\Lambda r^2}{3}\biggr)^{-1}dr^2\\
+r^2(d\theta^2+sin^2\theta d\phi^2),
\end{eqnarray} 
\noindent
where $ds$ is the line element of the 4-dimensional spacetime, and $t, r, \theta, \phi$ are the spherical polar coordinates, centered on the body.

The deviation from the Minkowski ``flat'' metric is minimal when

\begin{eqnarray}
r=R_V=\biggl(\frac{3GM}{\Lambda c^2}\biggr)^{1/3}.
\end{eqnarray} 
\noindent
The surface defined by this radial distance from the center is called the zero gravity surface, and it is the weak field limit of General Relativity in the presence of the cosmological constant. It becomes significant in the smallest scales of cosmological expansion, such as in the Hubble flow around the Local Group of galaxies. J\"arnefelt (1933) was the first to derive Eq. 1.4 and the zero gravity surface is now understood to appear within the ``vacuole'' introduced by Einstein and Straus (1945, 1946) to describe the environments of bound mass concentrations in expanding space. This weak field limit is our first topic of testing General Relativity which we  describe in detail.
 
The weak field is often also defined by (Psaltis 2008)
\begin{eqnarray}
v/c\ll1,& r_s/r\ll1
\end{eqnarray} 
while in case of a strong field we have
\begin{eqnarray}
v/c\sim1,&        r_s/r\sim1.
\end{eqnarray} 
Here $v$ is the orbital speed or other characteristic velocity, $r$ is the orbital radius or other characteristic length scale in the system and $r_s$ is the Schwarzschild radius of the primary body of mass $M$
\begin{eqnarray}
r_s=\frac{2 GM}{c^2}.
\end{eqnarray}

In the first weak field category belong e.g. the precession of Mercury's orbit, bending of light near the Sun, precession of a binary pulsar orbit,  gravitational radiation from the binary pulsar, relativistic geodetic precession and the precession due to the relativistic Lense-Thirring effect. In the latter strong field category we have at present only the binary black hole system OJ287, and hopefully in future other merging black hole binaries. The strong field limit in OJ287 is our second detailed topic of testing General Relativity.

Before these detailed discussions, we give a historical review of tests of General Relativity. Historically, the first test case supporting General Relativity was the non-Newtonian precession of the major axis of the planet Mercury.  In Newtonian theory,  a small object orbiting a concentrated spherically symmetric central body should retrace the same elliptical path repeatedly.  In case of Mercury, the major axis of its elliptical orbit precesses slowly. The precession is mostly due to gravitation of other planets. However, the mathematician Urbain Le Verrier of Paris Observatory concluded, after the influence of other planets have been deducted, that there was a remaining  observed shift of $38^{''}$/century in the perihelion of Mercury (Le Verrier 1859). He ascribed this to an unknown planet inside Mercury's orbit.  This presumed planet was never found despite extensive searches.  The United States Naval Observatory's Simon Newcomb (1895) recalculated the observed perihelion shift and obtained $41^{''}\pm2^{''}$/century (also see Hall 1894, Shapiro et al. 1976). 

 With General Relativity Einstein (1915) derived  an additional precession term to be added to the Newtonian precession.  Einstein's calculation gave a ``post-Newtonian'' (PN) force law which is almost but not quite the inverse square law  proposed by Newton.  For the post-Newtonian case, in one orbit, the ellipses traced by the two bodies orbiting each other precess through an angle given by

\begin{eqnarray}
\Delta\phi=\frac{6\pi GM}{c^2 a (1-e^2)}  
=3\pi\frac{r_s}{a(1-e^2)},
\end{eqnarray}
\noindent 
where $M$ is the total mass of the two bodies, $a$ is the semi-major axis of the relative motion and $e$ is the eccentricity. When one of the bodies is much greater than the other, as in the Sun -- Mercury system, the latter equality is useful since it scales the result to the Schwarzschild radius of the primary body.

To calculate a numerical value for the relativistic perihelion shift $\Delta\phi$ for Mercury, substitute the Schwarzschild radius for the Sun, $r_s = 2.96$ km, the semi-major axis of Mercury's orbit $a = 57.91\times10^6$ km, and its $e = 0.2056$ into Equation (1.8).  Multiply the result by 206,265 to go from radians to arcsec and by 415, the number of revolutions per 100 years to obtain
\begin{eqnarray}
\Delta\phi=43^{''}
\end{eqnarray} 
\noindent  
per century. The  agreement found in this early test of General Relativity gave great confidence in the theory. Today's accuracy of testing General Relativity in the Solar System is at a level of $0.01\%$. Here the characteristic velocity is $v/c\sim10^{-4}$, and we are dealing with weak fields.

Another more extreme and recent test of General Relativity is the radio pulsar PSR 1913+16, a smaller member of a binary system of two neutron stars about 1/60 as far from one another as Mercury is from the Sun. Using the semi-major axis of the binary $a \sim 1.95\times10^6$ km, the total mass $M \sim 2.83 M_{\odot}$ and eccentricity $e \sim 0.617$, we find that in one year (1131 orbital revolutions) the periastron advances

\begin{eqnarray}
\Delta\phi\sim 4.2^{\circ}.
\end{eqnarray} 

The orbital parameters result in a prediction for the rate of loss of the orbital energy by gravitational radiation. It matches observations with the accuracy of $0.2\%$ (Hulse and Taylor 1975, Taylor and Weisberg 1989, Damour and Taylor 1991, Hulse 1994). In binary pulsars, where the most interesting object is the double pulsar PRS J0737-3039 with the perihelion advance of $16.9^{\circ}$ per year (Lyne et al. 2004, Kramer et al. 2006), we have $v/c\sim2\times10^{-3}$. Here we are also testing General Relativity in weak fields, but including gravitational radiation.

The dragging of space around rotating bodies in General Relativity was proposed by Austrian physicists Joseph Lense and Hans Thirring (1918). By 2004 the Lense-Thirring effect was measured in space surrounding the rotating Earth. Following the motions of two Earth satellites LAGEOS I and II, a team lead by Ignazio Cuifolini of University of Lecce, Italy, and Erricos Pavlis of University of Maryland, found that the planes of the orbits of the satellites have shifted by about two meters per year in the direction of the Earth's rotation due to dragging of space around the Earth (Cuifolini and Pavlis 2004). The result is in agreement with the prediction of Lense and Thirring within the $10\%$ accuracy of the experiment. The satellite Gravity Probe B, designed for the measurement of space dragging confirmed these results in 2007 (Everitt et al. 2011). The Gravity Probe B Lens-Thirring value  is $0.04^{''}$/yr in contrast to the much larger $6.6^{''}$/yr for the geodetic precession of its 7,027 km, 97.65 minute orbit around the Earth. 

Relativistic spin-orbit coupling may cause the PSR 1913+16 pulsar's spin axis to precess (Damour and Ruffini 1974; Barker and O'Connell 1975a,b, Weisberg and Taylor 2005, Weisberg et al. 2010). This precession can result in changes in pulse shape as the pulsar-observer geometry changes. Under the assumption that relativistic precession is occurring, these changes have been used to model the two-dimensional structure of the pulsar beam.  

Another early convincing test of the General Theory of Relativity is the bending of light rays which pass close to the Sun.  Consider a photon passing by a mass $M$ with $b$ as the minimum distance between the photon and the mass $M$.  
The total deflection from a straight line path is (Einstein 1911)

\begin{eqnarray}
\Delta\phi=\frac{4 GM}{c^2 b}=2\frac{r_s}{b}
\end{eqnarray} 
\noindent
radians. For a light ray passing the surface of the Sun, $r_s = 2.96$ km and $b = 6.96\times10^5$ km. General Relativity gives
\begin{eqnarray}
\Delta\phi=1.75^{''},
\end{eqnarray} 
in good agreement with observations. Today better measurements using radio sources agree with Einstein's theory within $1\%$ accuracy. A classical Newtonian calculation only gives half the relativistic value.   Bending of light provided the second early convincing test of the General Relativity theory  (Eddington 1919, Dyson et al. 1920).  Via the deflection, stars close to the edge of the Sun appear to shift radially outward from the center of the Sun during a solar eclipse compared to a photograph of the same area six months earlier.

In a remarkable extension of the classical test of relativity beyond using the Sun, deflection of light from background sources has been seen around massive objects resulting  in gravitational lensing, multiple images or strong distortions of the sources.  The apparent source flux may also change.  Multiple images of the quasar Q0957+561 were detected in 1979 by a team led by Dennis Walsh of the University of Manchester (Walsh et al. 1979) using the 2.1 meter telescope of Kitt Peak National Observatory.  Nowadays gravitational lenses are detected frequently, and are used in astrophysical studies, in particular in estimating the relative amounts of dark matter in clusters of galaxies and matter's importance relative to dark energy in the universe.

One of the phenomena related to the elasticity of space is gravitational waves, small changes in the curvature of space which propagate in space with speed of light. At the moment, evidence for gravitational waves is indirect. The binary neutron star system PSR 1913+16 appears to emit gravitational waves. Observations show that the binary system does lose energy which cannot be explained in other ways beside gravitational wave emission. The loss rate of energy matches rather well what is expected in the General Relativity. This coincidence is usually taken as proof that gravitational waves do exist, even though the radiation from PSR 1913+16 is not presently directly measurable by gravitational wave antennas (Weisberg and Taylor 2005).

A promising case for both direct detection of gravitational radiation  and  the study of relativistic spin-orbit coupling is the binary black hole system of quasar OJ287 to be discussed later (Valtonen and Lehto 1997). Here one of the members is more massive than a star by a factor of $10^{10}$. Thus gravitational wave emission from this source should be very much more powerful than from the neutron stars of PSR 1913+16. The next generation of gravitational wave antennas should be able to directly confirm the emission of gravitational waves (Sun et al. 2011, Liu et al. 2012).

 The curvature of spacetime around a rotating black hole was first calculated by the New Zealand mathematician Roy Kerr (1963).  By conservation of angular momentum, a black hole arising from a rotating body must also rotate. The rotation of the black hole influences the surrounding spacetime even well beyond the black hole's Schwarzschild radius.

The quadrupole moment of the spinning primary black hole in OJ287 has a measurable effect on the orbit of the secondary. In OJ287, $v/c\sim0.25$. Thus here we are carrying out strong field tests of General Relativity. For example, it has been shown that the loss of orbital energy from the system agrees with General Relativity with the accuracy of $2\%$ (Valtonen et al. 2010b). More importatly, we may test the no hair theorems of black holes (Israel 1967, 1968, Carter 1970, Hawking 1971, 1972, see also Misner, Thorne and Wheeler 1973) for the first time. They relate the spin and the mass of a black hole to its quadrupole moment in a unique way (see Section 5).

Based on General Relativity, Einstein (1917) proposed a model for a curved finite (but still boundless)  universe. The cosmological constant, $\Lambda$, was specified so as to produce a static universe with no origin in time.  The evolving models of the universe, standard today, were derived by the Russian Alexander Friedmann in two papers in 1922 and 1924. These papers,  a turning point in the study of cosmology, remained almost unnoticed.  In 1927 Belgian astronomer Georges Lema\^itre rediscovered these models, now known as Friedmann universes.

Friedmann found a solution of the General Relativity equations which is more general than Einstein's solution. The Friedmann solution includes the Einstein static solution as a special case, where the cosmological constant is non-zero and related to the matter density so as to attain the gravity-antigravity balance. However, generally, the solution puts no restrictions on the cosmological constant.  It might be zero or nonzero, positive or negative, related or unrelated to the matter density.  The solution depends explicitly on time. The universe is not static: it expands or contracts as a whole.

Friedmann preferred expansion over contraction citing observational evidence supporting his choice in Slipher's data of galaxies that are moving away from us.  Friedmann died in 1925 at the age of 37 before the Hubble discovery of the redshift distance relation (Hubble 1929).  We can thus regard the discovery of the Hubble expansion as a predictive  verification of General Relativity.

After the discovery of the expansion of the universe, and other evidence for an origin in time, the so called Big Bang, it became obvious that $\Lambda$ could not be as big as Einstein had calculated for his static universe model. It became common to assume that $\Lambda=0$.  The ideas changed in the late 1990s when it became possible to use extremely luminous standard candles, supernovae of type Ia, to estimate distances of galaxies whose redshifts $z$ are comparable to unity.
In 1998-99 two groups discovered the non-zero $\Lambda$, usually interpreted as an indication of cosmic vacuum or dark energy (Riess et al. 1998 and Perlmutter et al. 1999).

In General Relativity $\Lambda$ is a constant, but it could be imagined in other theories that $\Lambda$ would depend on cosmic time. This is something that we can test at the weak field limit of General Relativity. Local studies of $\Lambda$, if they show that it has the same value which is observed using distant supernovae, can in principle exclude many alternative models. This will be discussed in the following sections.

\section{Weak limit: concept of zero gravity surface}

In recent years it has become customary to move the $\Lambda$ term from the left hand side of Equation (1.1) to the right hand side. Then it may be viewed as a contribution to the energy momentum tensor. The corresponding density is called the vacuum density $\rho_V$, with the equation of state
\begin{eqnarray}
\rho_V = -P/c^2,
\end{eqnarray}
where $P$ is the vacuum pressure. The value of the vacuum density is related to $\Lambda$ by
\begin{eqnarray}
\Lambda=\frac{8\pi G}{c^2}\rho_V.
\end{eqnarray}
Instead of $\Lambda$ it is common to state its normalized value
\begin{eqnarray}
\Omega_{\Lambda}=\frac{c^2\Lambda}{3H_{0}^2}
\end{eqnarray}
which is a dimensionless number of the order of unity. $H_0$ is the Hubble constant. From the cosmological recession of distant galaxies using Ia supernovae, the analysis of the microwave background radiation (CMB) and by many other ways (see Section 5), we find
\begin{eqnarray}
\rho_V \approx 7 \times 10^{-30} g cm^{-3},
\Omega_{\Lambda} \approx 0.73.
\end{eqnarray}

According to General Relativity, gravity depends on pressure as well as density: the effective gravitating density
\begin{eqnarray}
\rho_{eff} = \rho + 3P/c^2.
\end{eqnarray}
It is negative for a vacuum: $\rho_{eff} = - 2\rho_V$, and this leads to repulsion (``antigravity''). Hence the study of the antigravity in our neighbourhood, and on short scales in general, is an important test of General Relativity and its concept of Einstein's $\Lambda$-term in weak gravity conditions.

One way to study the local dark energy is via an outflow model (Chernin 2001; Chernin et al. 2006; Byrd et al. 2011)
which describes 
expansion flows around local masses.  It was motivated by
the observed picture of the Local Group
with outflowing dwarf galaxies around it
(van den Bergh 1999; Karachentsev et al. 2009).
 The model treats the dwarfs as "test
particles" moving in the force field produced by the
gravitating mass of the group and the possible dark energy
background. 
A static and spherically symmetric gravitational potential is
a rather good approximation despite of the binary structure of the Local Group (Chernin et al. 2009).

\subsection{A gravitating system within dark energy and the zero-gravity radius}

Soon after the discovery of the universal acceleration from observations of distant supernovae, researchers returned to the old question of J\"arnefelt, Einstein and Straus about what happens to the spacetime around a mass concentration in an expanding universe and asked: at what distance from the Local Group do the gravity of its mass and the antigravity of the dark energy balance each other (Chernin 2001, Baryshev et al. 2001)? 

Treating a mass concentration as a point mass $M$ on the background of the antigraviting dark energy, its gravity produces the radial force
$-GM/r^2$,
where $r$ is the distance from the group barycenter. The antigravity of the vacuum produces the radial force
$G2\rho_V(4\pi/3)r^3/r^2 = (8\pi/3)\rho_V r$.
Here $-2\rho_V$ is the effective gravitating density of vacuum. Then the radial component of motion in this gravity/antigravity force field obeys the Newtonian equation

\begin{equation}
\ddot r = -GM/r^2 + r/A^2\, ,
\label{newton_eq}
\end{equation}
where $r$ is the distance of a particle to the barycenter of the mass concentration. The constant 

\begin{equation}
A = \left[(8\pi G/3)\rho_V\right]^{-1/2}
\label{vacuum_time}
\end{equation}
is the characteristic vacuum time and has the value 
$= 5 \times 10^{17}$ s (or 16 Gyr) for $\rho_V = 7 \times 10^{-30}$ g cm$^{-3}$.
 
Equation (2.6) shows that the gravity force ($\propto 1/r^2$) dominates the antigravity force ($\propto r$)
at short distances. At the "zero-gravity distance" 

\begin{equation}
R_V = \left(GMA^2 \right)^{1/3} = \left[(3/8\pi)(M/\rho_V)\right]
\label{zero_grav_distance}
\end{equation}
the gravity and antigravity balance each other, and the acceleration is zero. At larger distances,
$r > R_V$, antigravity dominates, and the acceleration is positive. For the Local Group mass 
$M \approx 4 \times 10^{12} M_{\odot}$
and the global dark energy density, the zero-gravity distance is $R_V \approx$ 1.6 Mpc.

Equation (1.4) shows that we arrive at the same concept also in full General Relativity.

If we try to calculate the zero-gravity radius around a point in a homogenous distribution, this radius will increase directly proportional to the radius of the considered matter sphere.  So, one cannot ascribe physical significance to the zero-gravity radius within a fully uniform universe. But this is not true once some structure appears in the universe.  A density enhancement does not appear alone, but together with the zero-gravity sphere around it. 

The zero-gravity sphere for a point mass $M$ has special significance in an expanding universe.  A light test particle at
$r > R_V$
experiences an acceleration outwards relative to the point mass.  If it has even a small recession velocity away from  $M$, it participates in an accelerated expansion.

For an isolated system of two identical point masses $M$, the zero-gravity distance, where the two masses have zero-acceleration relative to their center-of-mass, is about $1.26 R_V$ (Teerikorpi et al. 2005).
  If separated by a larger distance, the masses will experience outward acceleration.
  
These examples illustrate the general result that in vacuum-dominated expanding regions perturbations do not grow.  They also lead one to consider the Einstein-Straus (1945) solution where any local region may be described as a spherical expanding  "vacuole" embedded within the uniform distribution of matter (Chernin et al. 2006). 

The metric inside the vacuole is static and
the zero-gravity radius and the (present-time) Einstein-Straus vacuole radius are simply related:

\begin{equation}
R_{ES} (t_0) = (2\rho_V / \rho_M)^{1/3} R_V \, .
\label{r_ES_r_V_ratio}
\end{equation}
For instance, for the ratio $\rho_V / \rho_M = 0.7/0.3$ one obtains $R_{ES} (t_0) = 1.67 R_V$. The Einstein-Straus vacuole can be seen as the volume from which gravitation has gathered the matter to form the central mass concentration.

\begin{figure}
\epsfig{file= 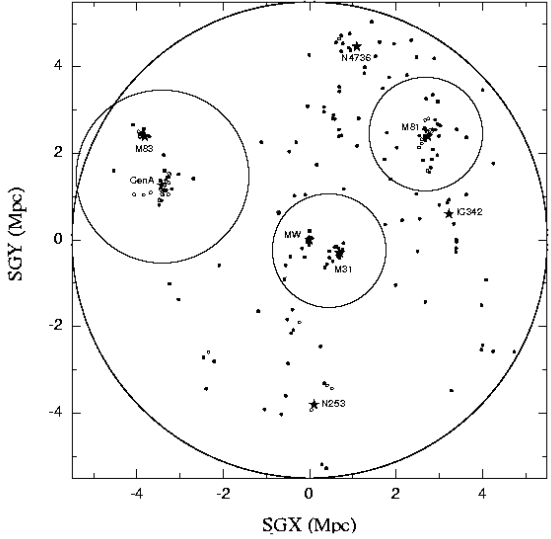, angle=0, width=8.0cm}
\caption{
Approximate zero-gravity spheres around the Local Group (at the center) and two nearby galaxy groups. The radii have been calculated using the masses
$2 \times 10^{12} M_{\odot}$
for the LG and the M81/M82 Group and
$7 \times 10^{12} M_{\odot}$
for the CenA/M83 Group (the underlying map presents the local environment up to about 5 Mpc as projected onto the supergalactic plane is from Karachentsev et al. 2003).}
\label{near_groups}
\end{figure}

Figure 1 shows the map of our local extragalactic environment up to about $5$ Mpc, together with the approximate zero-gravity spheres drawn around the Local Group and two nearby galaxy groups.  The spheres do not intersect.  This suggests that the groups are presently receding from each other with acceleration.


Using Eq. 2.8 one may calculate typical zero-gravity radii for different astronomical systems, for the standard value of $\rho_V$.  For stars, star clusters, galaxies and tight binary galaxies the zero-gravity radius is much larger than the size of the system which is located deep in the gravity-dominated region. For galaxy groups and clusters, $R_V$ is near or within the region where the outflow of galaxies begins to be observed (about 1.6 Mpc for the Local Group).  It is especially on such scales where the system and its close neighborhood could shed light on the local density of dark energy.  One is led to ask what happens to the test particles (dwarf galaxies) that have left the central region of the system? 

\subsection{Dynamical structure of a gravitating system within dark energy}

The particles move radially practically as predicted by the Newtonian equation of motion (Eq. 2.6), where the forces are the gravity of the central mass and the antigravity of the dark energy. The first integral of this equation expresses the mechanical energy conservation:
\begin{equation}
\frac{1}{2}\dot r^2 = E - U(r)\, ,
\label{first_integral}
\end{equation}
where $E$ is the total mechanical energy of a particle (per unit mass) and $U(r)$ is the potential energy
\begin{equation}
U(r) = - \frac{GM}{r} - \frac{1}{2}\left(\frac{r}{A} \right)^2. 
\label{first_integral}
\end{equation}
Because of the vacuum, the trajectories with $E< 0$ are not necessarily finite.  Such behavior of the potential has a clear analogue in General Relativity applied to the same problem.

The total energy of a particle that has escaped from the gravity potential well
of the system must exceed the maximal value of the potential:
\begin{equation}
E > U_{\mathrm{max}} = -\frac{3}{2} \frac{GM}{R_{\mathrm{V}}}\, .
\end{equation}

 It is convenient to normalize the equations
to the zero-gravity distance $R_{\rm V}$ and consider the
 Hubble diagram with normalized distance and velocity: $x = r/ R_{\rm V}$ and
 $y = (V/H_{\rm V}) R_{\rm V}$ where $V$ is the radial velocity and $H_{\rm V}$ is a constant to be discussed below
(Teerikorpi et al. 2008).
 Then radially moving test particles will move along curves which depend only on the constant total
 mechanical energy $E$ of the particle:
\begin{equation}
 y = x(1 + 2x^{-3}- 2\alpha x^{-2})^{1/2} \, .
\end{equation}  
 Here $\alpha$ parameterizes the energy, so
that $E = - \alpha GM/R_{\rm V}$.
Each curve has a velocity minimum at
$x =1$, i.e. at $r = R_ {\rm V}$.

 The
energy with $\alpha = 3/2$ is the minimum energy which still allows a particle initially
below $x = 1$ to reach this zero-gravity border (and if the energy is slightly larger) to continue
to the vacuum-dominated region $x > 1$, where it starts accelerating.
In the ideal case one does not expect particles with $x > 1$ below this minimum velocity curve.

\begin{figure*}
\centering
\epsfig{file=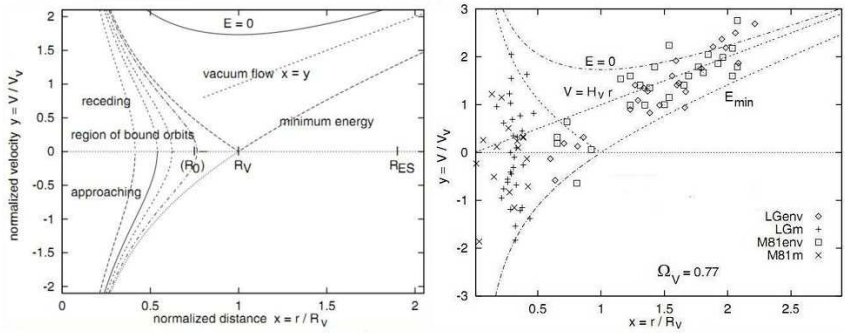, angle=0, width=12cm}
\caption{
Left panel: Different regions in the normalized Hubble diagram around a point mass in the vacuum. In the region of bound orbits a dwarf galaxy cannot move into the vacuum flow region unless it receives extra energy as a result of an interaction with other galaxies.
Right panel: The normalized Hubble diagram for the galaxies in the environments of the LG and M 81 groups, for the near standard vacuum density.  The velocity-distance relation for the vacuum flow is shown.  The curve for the lower limit velocity is given below and above $x =1$; below $x =1$ its negative counterpart is shown.  The members of the groups,  refered to by ``m'' in the symbol list, are found within these curves.
 (Adapted from Teerikorpi et al. 2008).}
\label{structure2_groups}
\end{figure*}

Figure 2 (left panel) shows different regions in the normalized Hubble diagram.  Below $r = R_V$, we have indicated the positive minimum velocity curve and its negative symmetric counterpart.  This region defines the bound group: a galaxy will not escape beyond $R_V$ unless it obtains sufficient energy from an interaction.

The diagonal line $x = y$ gives the "vacuum" flow with the Hubble constant $H_V$, when dark energy is fully dominating.
It is asymptotically approached by the out-flying particles beyond $x =1$.
This limit is described by de Sitter's static solution which has the metric of Eq. (1.2) with
$M=0$.
The spacetime of de Sitter's solution is determined by the vacuum alone, which is always static. It leads to the linear velocity--distance law, $V = r / A = H_V r$, with the constant expansion rate $H_V = (8\pi G\rho_V/3)^{1/2}$:
\begin{equation}
H_V = 61 \times \left(\frac{\rho_V}{ 7 \times 10^{-30} g/cm^3} \right)^{1/2}  km/s/Mpc.
\end{equation}
The normalized vacuum energy density depends on $H_{\rm V}$: $\Omega_{\Lambda}=H_{\rm V }^2/H_0^2$. The vacuum Hubble time $T_V = 1/H_{\rm V}$ exceeds the global Hubble time ($=1/H_0$)  by the factor
$(1+\rho_M / \rho_V)^{1/2} = (\Omega_{\Lambda})^{-1/2}$
for a flat universe.  In the standard model 
$T_V = 16 \times 10^9$ yr and the age of the universe ($13.7 \times 10^9$ yr) is about $0.85 T_V$.

Figure 2 (right panel) shows a combined normalized Hubble diagram for the Local Group, the nearby M81 group, and their environments.  The M81 group is at a distance of about 4 Mpc.   We have used in this diagram the mass $2 \times 10^{12} M_{\odot}$ for both groups. The energy condition to overcome the potential well ($ E > -3/2 GM/R_{\rm V}$) is not violated in the relevant range $x = r/R_{\rm V} >1$.

\section{On local detection of dark energy: the Local Group}

One expects inward acceleration at distances $r < R_V$ and outward acceleration at distances $r > R_V$
within the region where the point-mass model is adequate.  
However, such accelerations in the nearby velocity field around the Local Goup
are very small (of the order of 0.001 cm/s/yr) and impossible to measure directly.
We also cannot follow a dwarf galaxy in its trajectory for millions of years in order to see the location of the minimum velocity which defines theSeta zero-gravity distance. 

The objects were likely expelled in the distant past within a rather narrow time interval (e.g., Chernin et al. 2004).  What we see now is a locus of points at different distances from the center and lying on different energy curves; they make the observed velocity-distance relation. 
 
\subsection{The present-day local Hubble flow}

The prediction for the present-day outflow of dwarf galaxies near the Local Group
(or other galaxy systems) depends on the mass of the group, the flight time  (< the age of the universe), and
the local dark energy density.

Peirani \& de Freitas Pacheco (2008) derived the velocity-distance relation using the Lema\^itre-Tolman model containing the cosmological constant, and compared this with the model for $\Lambda = 0$, using the central
mass and the Hubble constant as free parameters.
Chernin et al. (2009) considered
the total energy for different outflow velocities at a fixed distance, and then
calculated the required time for the test particle to fly from near the group's center up to
this distance. The locus of the points corresponding to a constant age ($\approx$ the age of
the Universe) gave the expected relation.

Here we show some results from a modification of the above methods which easily allows one to vary the values of the relevant parameters and to generalize
the model in various ways (Saarinen and Teerikorpi 2014). One generates particles close to the center of mass of the group and
gives them a distribution of speeds. Then they are allowed to fly along the radial direction for a time $T$,
and their distances from the center and velocities are noted. The flight time $T$, during which the integration of the equations of
motion is performed, is at most the age of the standard
Universe, 13.7 Gyr. 

When comparing the standard $\Lambda$ CDM model, with its constant dark energy density on all
scales, it is relevant to consider the "Swiss cheese (SwCh) model", where the Universe has the same age as the standard model, but where the dark energy density is zero on small scales.
This model could correspond to the case where dark energy (or analogous effects) operates on large scales only. 

We plot some results together with the data on the local outflow around the Local Group
from the Hubble diagram as  derived  by Karachentsev et al. (2009) from   HST observations. 
The largest available distance is 3 Mpc.

\begin{figure*}
\centering
\epsfig{file=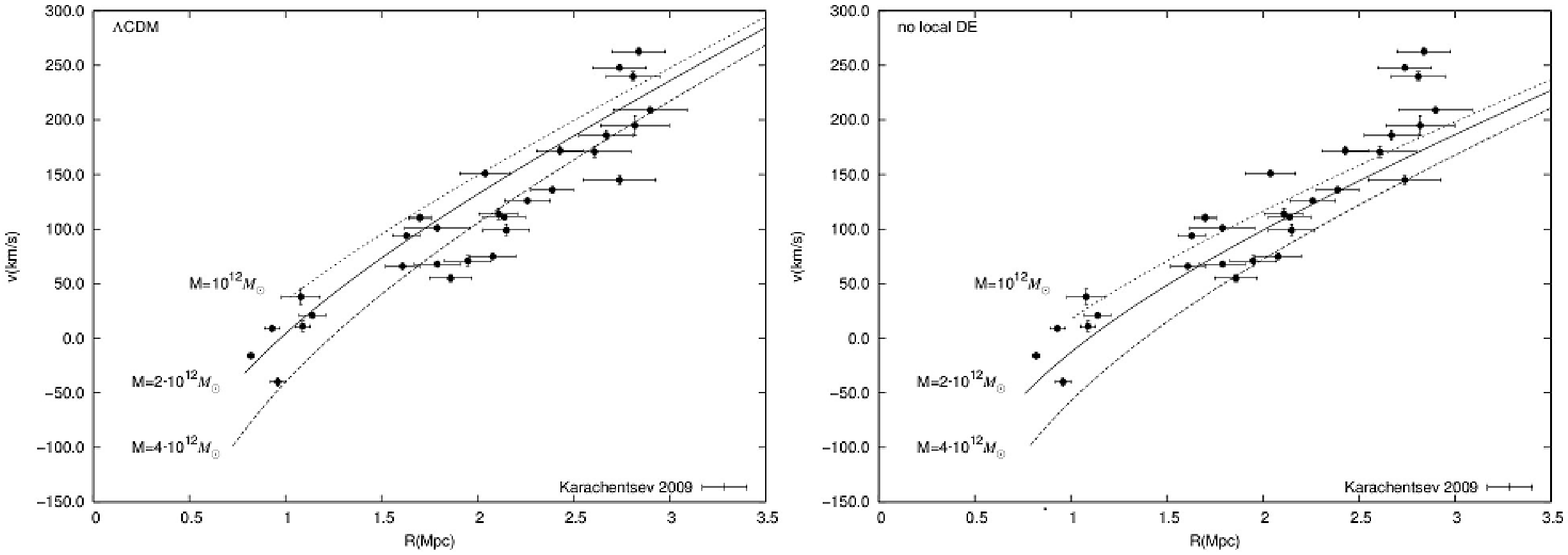, angle=0, width=12cm}
\caption{Left  panel: The location of test particles as injected from the mass centre (curves) after the flight time $T_0 = 13.7$ Gyr,
for different masses ($ 1 \times 10^{12} M_{\odot}$, $ 2 \times 10^{12} M_{\odot}$, $ 4 \times 10^{12} M_{\odot}$).
Here the standard model is used (local dark energy = global dark energy).
Right panel: The location of test particles after the flight time $T_0 = 13.7$ Gyr,
for different masses.
Here the Swiss cheese model is used (no local dark energy).
The data points are for the Local Group. 
(Credit. J. Saarinen)}
\label{fig_multimass_de_swch137}
\end{figure*}

Figure 3 shows the predicted distance--velocity curves for different masses in the standard model
(left panel) and for the SwCh model (right panel). It is seen that
the location of the curve depends rather strongly on the adopted mass. For instance, the SwCh curve with $ M = 2 \times 10^{12} M_{\odot}$ fits the LG data along the whole distance range. 
The $\Lambda$CDM model requires $ M \approx 4 \times 10^{12} M_{\odot}$ for a good
fit beyond the zero-gravity distance. 

Figure 3 shows the cases when the particles were ejected just after the Big Bang. 
This case would correspond to a classical situation where the outflow around the central group is "primordial".
In practice, the age of the group and the outflowing dwarf galaxies must be less than the age of the Universe, and the origin of the outflow may be due to early interactions within the system, making galaxies escape from it (e.g., Valtonen et al. 1993, Byrd et al. 1994, Chernin et al. 2004). The dark energy antigravity enhances the escape probability because it makes the
particle potential energy barrier lower than in the presence of gravity only.


The calculations show clear differences between the two cases in the sense expected: the $\Lambda$CDM curves are above the no-local-dark-energy curves and steeper, making the zero-velocity distance longer in the latter case, as already noted  by Peirani \& de Freitas Pacheco (2008) and Chernin et al.
(2009). However, in practice, the difference may be difficult to detect. First, the observed distance -- velocity relation is rather scattered. Second, the independently known mass of the Local Group is uncertain. Thirdly, the model of the galaxy group and its evolution contains uncertain elements, including the exact ejection time.

\subsection{Mass, dark energy density and the "lost gravity" effect}

A general conclusion from the outflow data is that a low-mass Local Group ($M \leq 2 \times 10^{12} M_{\odot}$) is associated with the case of no local dark energy, while a local density near the global dark energy density requires a higher mass, 
$M \approx 4 \times 10^{12} M_{\odot}$. This mutual dependence between the assumed mass and the derived dark energy density is typical for various local dynamical tests.

We may estimate conservative limits of local dark energy density $\rho_{loc}$ as follows. If the value of $R_{\rm V}$
 were known from the velocity-distance diagram
and the mass $M$ of the group is independently measured, the
dark energy density may be estimated in the outflow region:
\begin{equation} \frac{\rho_{loc}}{\rho_{V}} = (\frac{M}{1.3 \times 10^{12}
M_{\odot}}) (\frac{1.3 {\rm Mpc}}{R_{\rm V}})^3.
\label{rh_M_R}
\end{equation}
In fact, attempts to probe dark energy with nearby outflows were
first made (e.g., Chernin et al. 2006, 2007; Teerikorpi et al. 2008) by using Equation (3.1).
Also here the derived dark energy density depends directly on the assumed mass in addition
to the strong inverse dependence on the used zero gravity distance. 

The size of the group is a strict lower limit to $R_V$, giving an upper limit to the local dark energy density $\rho_{\mathrm loc}$,
for a fixed group mass. Hence, for the Local Group $R_V > 1 $ Mpc leads to $\rho_{loc} < 2.2\rho_{V}$, for
the mass $M = 2 \times 10^{12} M_{\odot}$.

An upper limit for $R_V$ would give the interesting lower limit to $\rho_{loc}$. One way to study $R_V$ would be to find the distance $R_{ES}$, the Einstein-Straus radius where the local outflow reaches the global Hubble rate of Eq. (2.14) (Teerikorpi and Chernin 2010).

If there is no local dark energy, the outflow reaches the global expansion rate only asymptotically in this idealized point mass model. However, assuming that $\rho_{loc} = \rho_V$, one may expect the local flow to intersect the global Hubble relation at a distance
$ R_{ES} = 1.7 R_V$. For instance, with the Local Group mass of $(2 - 4) \times 10^{12} M_{\odot}$  and using the global
dark energy density, one calculates $R_{ES} = 1.7 R_V = 2.2 - 2.6$ Mpc. This range is indeed near the distance where the local expansion reaches the global rate (Figure 3; see also Karachentsev et al. 2009).

Starting with the illustrative value $R_{ES} \approx 2.6$ Mpc, one may estimate that $R_V \approx 1.5 $ Mpc (Teerikorpi and Chernin 2010) and a lower limit for
the dark energy density around the Local Group would be  $\rho_{loc}/\rho_V \geq 0.4$ for the mass of $2\times 10^{12}
M_{\odot}$. The limit is directly proportional to the adopted mass value.

In view of the uncertainties in using the outflow kinematics only, one should use in concert other
independent methods for putting limits on the value of the local dark energy density. Such include the Kahn-Woltjer
method and the virial theorem,  both of which can be modified to take into account the "lost gravity" effect of  dark energy ( Chernin et al. 2009).

\begin{figure}
\epsfig{file=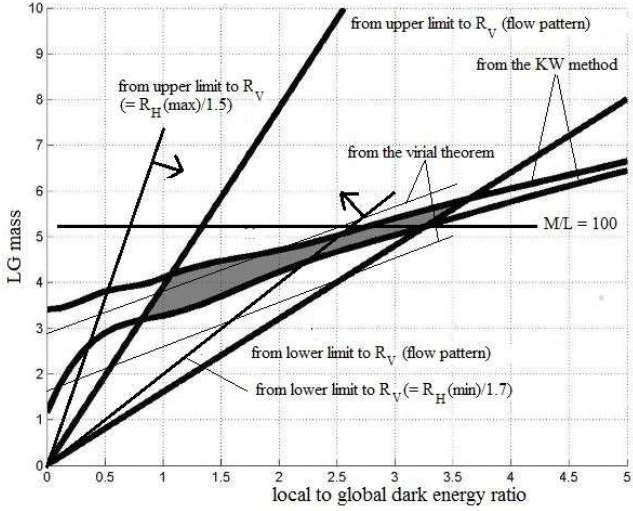, angle=0, width=9.0cm}
\caption{Results on the very local density of dark energy from the Local Group and its near environment (adapted from Chernin et al.
(2009). 
}
\label{LG_combined}
\end{figure}

Kahn and Woltjer (1959) used a simple one-dimensional two body problem to describe the relative motion of the Milky Way and M31 galaxies.  The motion of the galaxies was described as a bound system. The (currently) observed values $r=0.7$ Mpc and $V = -120$ km/s  lead to a limit for the estimated binary mass:
$M > 1 \times 10^{12} M_{\odot}$.
The mass is
$M \approx 4.5 \times 10^{12} M_{\odot}$,
if the maximum separation was about 4.4 Gyr ago, corresponding to 13.2 Gyr as the time since the two galaxies started to separate from each other.

With a minimal modification of the original method,  the first integral of the equation of motion (eq. (2.10)), in the presence of dark
energy background, becomes:
\begin{equation}
\frac{1}{2} V^2= \frac{GM}{r} + \left (\frac{G4\pi}{3}\right )\rho_V r^2 + E.
\end{equation}
Now the total energy for a bound system embedded in the dark energy background must be smaller than an upper limit which depends on the mass and the dark energy density:
\begin{equation}
E < -\frac{3}{2}GM^{3/2}\left [\left (\frac{8\pi}{3}\right )\rho_V \right ]^{1/3}.
\end{equation}
The limiting value corresponds to the case where the distance between the component galaxies could just reach the zero-gravity distance. 
Now the lower mass limit increases to
$M > 3.2 \times 10^{12} M_{\odot}$.
  Also the timing argument leads to an increased mass, 
$M \approx 5.3 \times 10^{12} M_{\odot}$ (Chernin et al. 2009; Binney \& Tremaine 2008).  On the other hand, dynamical activity during the formation and settling down of the Local Group tends to reduce ``timing'' mass by a large factor (Valtonen et al. 1993). Though the uncertainty due to the various imperfectly known factors is rather large, the method suggests
 \begin{equation}
M \sim 4 \times 10^{12} M_{\odot}
\end{equation}
 for the Local Group. A local volume cosmological simulation with this mass value agrees nicely with observations (Garrison-Kimmel et al. 2013).

The classical virial theorem is a well-known way to determine the mass of a quasi-stationary gravitationally bound many-particle system.  For a system within dark energy, the virial theorem needs an extra term due to the contribution of the particle-dark energy interaction to the total potential energy (Forman 1970; Jackson 1970). When positive, the cosmological constant leads to a correction upwards for the mass estimates (Chernin et al. 2009) and one can show that the modified virial theorem should be written, in terms of the total mass $M$, a characteristic velocity $V$ and a characteristic sizes $R$, as:
\begin{equation}
M = V^2 \frac{R}{G} + \frac{8\pi}{3}\rho_V R^3 \, ,
\end{equation}
where the second term on the right hand side is the correction due to the dark energy. It is interestingly equal to the value of the effective (anti)gravitating mass of dark energy within the sphere of radius $R$.  It is a measure of the lost-gravity effect, which
can be significant in galaxy groups. In the Local Group the correction term contributes about 30 percent of the total mass.

Chernin et al. (2009) used the
modified Kahn-Woltjer method together with the outflow data
to derive the mass of the Local Group and the local
dark energy density, resulting in $M = (3-6) \times
10^{12} M_{\odot}$ and  $ \rho_{loc}/\rho_{V} = 0.8-3.8$.
The virial estimator which uses the mass and
the velocity dispersion within the Local Group
(Chernin et al. 2012) gave $\rho_{loc}/\rho_{V} = 0.7-2.8$.

We present various Local Group results  in Fig.4 where the horizontal axis gives the assumed local-to-global
dark energy density and the vertical axis is the mass of the Local Group. A simpler version of this diagram appeared in Chernin et al.
(2009), where the admissible range from the modified Kahn-Woltjer --method was shown, together with the limiting straight lines
corresponding to the upper and lower limit of $R_V$ as estimated from the appearance of the outflow pattern. These
defined the darkened area as the possible range of the mass and the local dark energy density. The present
Fig.4 includes additional constraints:

1) The range of the virial mass based on van den Bergh's (1999) result $M_{\rm LG} = 2.3 \pm 0.6 M_{\odot}$ and using the correction term in the appendix to Chernin et al. (2009).

2) The location of the mass-to-luminosity ratio $M/L = 100$ (an upper limit for small groups) as the horizontal line, based on van den Bergh's (1999) result that $M_{\rm LG} = 2.3 \times 10^{12} M_{\odot}$ corresponds to $M/L = 44$, and

3) The lines corresponding to upper and lower limits to $R_V$ as inferred from the distance range 2.2 -- 3 Mpc where the local flow reaches the global Hubble rate and using the calculations in Teerikorpi \& Chernin (2010). These limits may be less subjective than the original thick lines which were based on visual inspection of the flow pattern.

We see that the different constraints from the Local Group put the local dark energy density into the range $ \rho_{loc}/\rho_{V} = 0.5-2.5$.

The preceding discussion did not make any use of the prior knowledge of the dwarf galaxy outflow. Dwarf galaxies are thrown out of the Local Group during its early assembly and later evolution. In cosmological N-body simulations where this process is seen, it is sometimes called a ``back splash''. The ``back splash'' follows the normal rules of dynamical ejection in a potential well (Valtonen and Karttunen 2006). The dwarf ejection times have a wide distribution, with a higher rate in the beginning. However, since the Local Group is dynamically young, as the two major galaxies have not yet completed even one full orbit, the dwarf ejection rate has not yet declined signicantly.

The distribution of the ejection speeds $P(V)$ is a steeply declining function of $V$. The probability that the escape velocity is in the interval $V, V+dV$ is
\begin{equation}
P dV \sim V^{-4.5} dV
\end{equation}
(Valtonen and Karttunen 2006, Eq. (11.33)). What it means that in practice all dwarfs cross the zero gravity surface with a speed which is very close to zero ($\sim40$ km/s). It has two consequences: the flow speed of dwarfs beyond the surface is very smooth, and the time spent by the dwarfs at the zero gravity sphere is long and consequently we find an accumulation of the flow at this boundary. This makes the identification of the zero gravity radius simple. For the Local group
\begin{equation}
R_{V} = 1.6 \pm 0.1 Mpc.
\end{equation}

The two main galaxies lose orbital energy from their relative motion every time a galaxy is ejected from the group. Therefore the Local Group timing argument (Kahn and Woltjer 1959), when applied to the isolated pair, gives necessarily an overestimate of their combined mass. The ejections shorten the major axis of the relative orbit, and thus less mass is required to close the orbit. The importance of this effect depends on the total mass of the ejected galaxies. In the scenarios calculated by Valtonen et al. (1993) the effect is $5\%$ in the universe of 14 billion yr in age.  A greater reduction in the ``timing mass'', up to $25\%$, comes from the possibility that the rotation speed of the Galaxy is greater than the standard 220 km/s. On the other hand, the lost gravity effect would tend to increase the ``timing mass'' by about $30\%$.

Therefore, all in all, the full N-body simulation model of Garrison-Kimmel et al. (2013), with Local Group mass of $M=4\times10^{12}$ solar mass, appears to be an acceptable model of the Local Group and the surrounding flows. With these values we get $\Omega_{\Lambda}=0.75$, the value given on the first line of  Table 1 (Section 5). The one standard deviation uncertainty may be estimated as $\pm10\%$. 

Another way to obtain the local $\Omega_{\Lambda}$, is to take the measured values of $H_V$ and $H_0$ from observations, and calculate
\begin{equation}
 \Omega_{\Lambda}=(\frac{H_V}{H_0})^2. 
\end{equation}
Even though both quantities on the right hand side have associated uncertainties, we get using $H_V=59$ km/s/Mpc and $H_0=70$ km/s/Mpc, an estimate $\Omega_{\Lambda}=0.71$ (Chernin 2013), again with the estimated one standard deviation uncertainty of $\pm10\%$. This value is given on the second line of Table 1.


\section{Dark energy in the Coma cluster of
galaxies}

In this Section, we extend our studies of the local dark energy
effects from groups of galaxies to clusters of galaxies
and address the Coma cluster considering it as the largest
regular, nearly spherically-symmetrical, quasi-stationary
gravitationally bound aggregation of dark matter and baryons
embedded in the uniform background of dark energy. Is antigravity
produced by dark energy significant in the volume of the cluster?
Does it affect the structure of the cluster? Can antigravity put
limits on the major gross parameters of the system? In a search
for answers to these questions, we will use and develop the
general considerations on the local dark energy given in the
sections above.

\subsection*{4.1.  Three masses of the cluster}

The mass of the Coma cluster was first measured by Zwicky
(1933,1937) decades ago. Using the virial theorem he found that it
was $3 \times 10^{14} M_{\odot}$ when normalized to
the presently adopted value of the Hubble constant $H_0 = 70$ km/s/Mpc
which is used here. Later The \& White (1986) found an order
of magnitude larger value, $2 \times 10^{15} M_{\odot}$, with a
modified version of the virial theorem. Hughes (1989, 1998)
obtained a similar value $(1-2) \times 10^{15} M_{\odot}$ with
X-ray data under the assumption that the hot intergalactic
gas in the cluster is in hydrostatic equilibrium. With a similar assumption,
Colless (2002) reports the mass $4.4 \times 10^{14} M_{\odot}$
inside the radius of 1.4 Mpc. A weak-lensing analysis gave the
mass of $2.6 \times 10^{15} M_{\odot}$ (Kubo et al. 2007) within
4.8 Mpc radius. Geller et al. (1999, 2011) examined the outskirts
of the cluster with the use of the caustic technique (Diaferio \&
Geller 1997, Diaferio 1999) and found the mass $2.4 \times 10^{15}
M_{\odot}$ within the 14 Mpc radius. Taken at face
value, it appears that the mass within 14 Mpc is
smaller than the mass within 4.8 Mpc. Most probably, this is due
to uncertainties in mass determination. Indeed, the $2\sigma$
error is $1.2 \times 10^{15} M_{\odot}$ in Geller's et al. (1999,
2011) data, and within this uncertainty, the result does not
contradict the small-radius data.

Also one should note that the action of the central binary in the Coma cluster has the effect of ejecting galaxies from the cluster. It leads to an overestimate of the cluster mass if virial theorem is used (Valtonen and Byrd 1979, Valtonen et al. 1985, Laine et al. 2004).

In each of the works mentioned here, the measured mass is treated
as the matter (dark matter and baryons) mass of the cluster at
various clustercentric distances. However, the presence of dark
energy in the volume of the cluster modifies this treatment,
since dark energy makes its specific contribution to the mass of
the system. This contribution is naturally measured by effective
gravitating mass of dark energy in the volume of the cluster at
various clustercentric distances $R$ (Eq. 3.5):
\begin{equation}
M_{V} (R) = \frac{4\pi}{3} \rho_{V eff} R^3 = - \frac{8\pi}{3}
\rho_{V} R^3 = - 0.85 \times 10^{12}  [\frac{R}{1 Mpc}]^3
M_{\odot}. 
\end{equation}
For the largest radius $R = 14$ Mpc we have:
\begin{equation}
 M_{V} = - 2.3 \times 10^{15} M_{\odot}, \;\;\; R = 14\; {\rm
Mpc}.
\end{equation}
The total gravitating mass within the radius $R$ is the sum
\begin{equation}
 M_G (R) = M_M (R) + M_{V}(R), 
\end{equation}
\noindent where $M_M (R)$ is the matter (dark matter and baryons)
mass of the cluster inside the same radius $R$. It is this mass
$M_G (R)$ that is only available for astronomical measurements via
gravity (with virial, lensing, caustic, etc. methods). Because of
this, we identify the gravitating mass $M_G(R)$ with the
observational masses quoted above for various clustrocentric
radii. In particular, the gravitating mass for the largest radius
$R = 14$ Mpc in the Coma cluster is this:
\begin{equation}
M_G(R) = M_V (R) + M_M (R) = 2.4 \times 10^{15} M_{\odot}.
\end{equation}
Then the matter mass $M_M$ at the same radius
\begin{equation}
 M_M(R) = M_G(R) - M_V(R) = 4.7 \times 10^{15} M_{\odot}.
\end{equation}

We see that the value of the matter mass $M_{M}$ at $R = 14$
Mpc obtained with the presence of dark energy is a
factor of (almost) two larger than that in the traditional
treatment. This implies that the antigravity effects of dark
energy are strong at large radii of the Coma cluster.

\subsection*{4.2.  Matter mass profile}

In the spherically symmetric approximation, the cluster matter mass $M_M(R)$
may be given in the form:
\begin{equation}
 M_M (R) = 4 \pi \int \rho(R) R^2 dR, 
\end{equation}

\noindent where $\rho (R)$ is the matter density at the radius
$R$. According to the widely used NFW density profile (Navarro et
al. 1997) 
\begin{equation}
 \rho = \frac{4\rho_s}{\frac{R}{R_s}(1 + \frac{R}{R_s})^2}, 
\end{equation}

\noindent where $\rho_s = \rho(R_s)$ and $R_s$ are constant
parameters. At small radii, $R << R_s$, the matter density goes to
infinity, $\rho \propto 1/R$ as $R$ goes to zero. At large
distances, $R >> R_s$, the density slope is $\rho \propto 1/R^3$.
With this profile, the matter mass profile is
\begin{equation}
 M_{M} (R) = 16 \pi \rho_s R_s^3 [\ln (1 + R/R_s) -
\frac{R/R_s}{1 + R/R_s}].  
\end{equation}

To find the parameters $\rho_s$ and $R_s$, we may use the
small-radii data (as quated above): $M_1 = 4.4 \times 10^{14}
M_{\odot}$ at $R_1 = 1.4$ Mpc, $M_2 = 2.6 \times 10^{15}
M_{\odot}$ at $R_2 = 4.8$ Mpc. At these radii, the gravitating
masses are practically equal to the matter masses there. The
values of $M_1, R_1$ and $M_2, R_2$ lead to two logarithmic
equations for the two parameters of the profile, which can easily
be solved: $ R_s = 4.7 \; \; {\rm Mpc}, \;\; \rho_s = 1.8 \times
10^{-28} \;\; {\rm g/cm}^3$. Then we find the matter mass within
$R = 14$ Mpc,
\begin{equation}
 M_{M} \simeq 8.7 \times 10^{15}\; M_{\odot}, 
\end{equation}

\noindent to be considerably larger (over 70\%) than  given by our
estimation above.

Another popular density profile (Hernquist 1990) is
\begin{equation}
 \rho(R) \propto \frac{1}{R (R + \alpha)^3}.  
\end{equation}

\noindent Its small-radius behavior is the same as in the NFW
profile: $\rho \rightarrow \infty$, as $R$ goes to zero. The slope
at large radii is different: $\rho \propto 1/R^4$. The
corresponding mass profile is
\begin{equation}
 M_{M} (R) = M_0 [\frac{R}{R + \alpha}]^2. 
\end{equation}
\noindent The  parameters $M_0$ and $\alpha$ can be found from the
same data as above on $M_1, R_1$ and $M_2, R_2$: $ M_0 = 1.4
\times 10^{16} \; M_{\odot}, \; \alpha = 6.4 \; {\rm Mpc}$,
giving another value for the mass within $14$ Mpc:
\begin{equation}
M_{M} = 6.6 \times 10^{15} M_{\odot}, \;\; R =  14 \; {\rm
Mpc}. 
\end{equation}

\noindent Now the difference from our estimated figure is about
40\%.

In a search for a most suitable mass profile for the Coma cluster,
we may try the following simple new relation:
\begin{equation}
 M_{M} (R) = M_* [\frac{R}{R + R_*}]^3.  
\end{equation}

\noindent This mass profile comes from the density profile:
\begin{equation}
 \rho (R) = \frac{3}{4\pi} M_* R_* (R + R_*)^{-4}.  
\end{equation}

\noindent The density goes to a constant as $R$ goes to zero; at
large radii, $\rho \propto 1/R^4$, as in Hernquist's profile.

The parameters $M_*$ and $R_*$ are found again from the data for
the radii of 1.4 and 4.8 Mpc: $ M_* = 8.7 \times 10^{15} \;
M_{\odot}, \; R_* = 2.4 \; {\rm Mpc}$. The new profile leads to
a lower matter mass at $14$ Mpc:
\begin{equation}
 M_{M} = 5.4 \times 10^{15} M_{\odot}, \;\;\;\ R = 14 Mpc, 
\end{equation}

\noindent which is equal to our estimate above within 15\%
accuracy.

\subsection*{4.3.  Upper limits and beyond}

It is obvious that a system of galaxies can be gravitationally
bound only if gravity dominates over antigravity in its volume. In
terms of the characteristic masses introduced above, this
condition may be given in the form:
\begin{equation}
M_{M}(R) \ge |M_{\rm V}(R)|. 
\end{equation}

\noindent The condition is naturally met in the interior of the
system. But the dark energy mass $|M_{\rm V}(R)|$ increases with
the radius as $R^3$, while the matter mass increases slower in all
the three versions of the matter mass profile $M_M (R)$ discussed
above. As a result, an absolute upper limit arises from this
condition to the total size and the total matter mass of the
cluster. For the new matter mass profile introduced above one has:
\begin{equation}
 R_{\rm max} = 20\; {\rm Mpc}, \;\;\; M_{\rm max}  = 6.2 \times
10^{15} M_{\odot}.
\end{equation}

These upper limits are consistent with the theory of large-scale
structure formation that claims the range $2 \times 10^{15} < M <
10^{16} M_{\odot}$ for the most massive bound objects in the
Universe (Holz \& Perlmutter 2012, Busha et al. 2005).

For comparison, the two traditional matter profiles mentioned
above lead to somewhat larger values of the size and mass:
\begin{equation}
 R_{\rm max} = 25\; {\rm Mpc}, \;\; M_{\rm max} = 1.5 \times
10^{16} M_{\odot} \; ({\rm NFW }),
\end{equation}
\begin{equation}
 R_{\rm max} = 22\; {\rm Mpc}, \;\; M_{\rm max} = 9.1 \times
10^{15} M_{\odot} \; ({\rm Hernquist}). 
\end{equation}

Generally, the limit condition $M_G(R_{\rm max})= |M_V(R_{\rm
max})|$ leads to the relation between the upper mass limit and the
upper size limit:
\begin{equation}
 R_{\rm max} = (\frac{3 M_{\max}}{8\pi \rho_V})^{1/3}. 
\end{equation}

\noindent The relation shows that the upper size limit is
identical to the zero-gravity radius introduced in Section 1: $R_{\rm
max} = R_{\rm V}$.

Studies of nearby systems like the Local Group and the Virgo and
Fornax clusters (Karachentsev et al. 2003, Chernin 2008, Chernin
et al. 2003, 2007, 2010, 2012a,b, 2013, Chernin 2013, Hartwick 2011) show that their
sizes are indeed near the corresponding zero-gravity radii, so
that each of the systems occupies practically all the volume of
its gravity-domination region ($R \le R_{\rm V}$). These examples
suggest that the Coma cluster may have the maximal possible size
and its total matter mass may be near the maximal possible value.
If this is the case, the mean matter density in the system is
expressed in terms of the dark energy density and does not depend
on the density profile (Merafina et al. 2012; Bisnovatyi-Kogan and
Chernin 2012, Bisnovatyi-Kogan and Merafina 2013):
\begin{equation}
 \langle \rho_{\rm M} \rangle = \frac{M_{M}}{\frac{4\pi}{3}
R_{\rm V}^3} = 2 \rho_{\rm V}. 
\end{equation}

Another theoretical prediction may be made about the environment of the
Coma cluster at the dark energy domination area $R > R_V$. 
Earlier studies have shown that outflows of galaxies exist around the
Local group and some other groups and clusters of galaxies. We may
assume that such an outflow may be observed around the Coma
cluster as well at the distance $R > R_V \simeq 20$ Mpc from its
center. The outflow is expected to have quasi-regular kinematical
structure with the nearly linear velocity-distance relation. The galaxies in the outflow have been ejected by dynamical interactions in the cluster center (Saarinen and Valtonen 1985, Byrd et al. 2007).

These theoretical predictions may be tested in current and future
astronomical observations of the Coma cluster and its environment. The next generation of ground based and space telescopes should be able to resolve individual stars in the galaxies of the Coma cluster, and allow us to construct a three-dimensional map of the cluster together with the radial velocities of each galaxy. We may then be able to separate the ``warm'' inflow of galaxies in the outskirts of the cluster from the ``cool'' outflow, and to determine the position of the zero-gravity radius experimentally.

\section{Testing the constancy of $\Lambda$}
The concept of dark energy (or more specifically Einstein's cosmological constant) has become a routine factor in global cosmology on Gpc scales. We have shown above that it is also relevant in the local extragalactic universe. In the local weak gravity conditions the antigravity effect of dark energy becomes measurable and has to be included in dynamical studies performed in galaxy group and cluster scales of a few Mpc. In general, extragalactic mass determinations should include a correction term due to the 'lost gravity' effect caused by dark energy.  Here we briefly mention a few other applications.

The Hubble outflow is one factor in explaining the well-known redshift anomaly in local galaxy group data via a selection effect (Byrd and Valtonen 1985, Valtonen and Byrd 1986, Niemi and Valtonen 2009). This asymmetry between redshifts and blueshifts of group members can be seen as a signature of local dark energy (Byrd et al. 2011).

The zero-gravity radius is an important quantity, which has existed (with roughly the same value) since the formation of a mass concentration in the expanding universe. It defines a natural upper limit to the size of a gravitationally bound system, allowing one to give an upper limit to the cosmic gravitating matter density in the form of galaxy systems. For the local universe this was discussed by Chernin et al. (2012b).

The measurement of the local value of the density of dark energy is naturally very important for our understanding of the nature of dark energy. Is it really constant on all scales and at all times, as suggested by the original concept of Einstein's cosmological constant? 

After the initial period of rediscovery of the cosmological constant $\Lambda$ in 1998-1999, there has been been a large amount of activity in trying to determine the exact value of the constant, and to study its possible dependence on redshift (Frieman et al. 2008, Blanchard 2010, Weinberg et al. 2013). As we stated before, in General Relativity $\Lambda$ is an absolute constant.

Tables 1-4 list a sample of $\Omega_{\Lambda}$ found in the literature from year 2005 onwards together with the one standard deviation error limits. The values cluster around $\Omega_{\Lambda}=0.73$ with a standard deviation of 0.044. This is somewhat smaller than the typical error in individual measurements. Only the cosmic microwave background (CMB) models give a significantly better accuracy, but in this case one may suspect hidden systematic errors due to foreground corrections (Whitbourn et al. 2014). Within the error bounds, $\Omega_{\Lambda}$ is constant over the redshift range that can be studied. Note that the CMB model values have been placed at redshift $ z=3$ since the higher redshift universe does not affect the derived $\Omega_{\Lambda}$ significantly. The $\Omega_{\Lambda}$ values obtained at the weak limit are the same as the cosmological determinations within errors.
\begin{table}[h] 
\caption{$\Omega_{\Lambda}$  determinations at different redshifts.\label{lambdatable}} 
\begin{tabular}{l|r|r|l}
\hline
  \hline
 Redshift&       $\Omega_{\Lambda}$&       Uncertainty&        Reference  \\
   \hline

0    &                   0.75 &                                  0.07 &                 this work\\
0    &                   0.71 &                                  0.07 &                 Chernin (2013)\\
0    &                   0.76 &                                  0.07 &                 Karachentsev et al. (2009)\\
0.005&                   0.78 &                                  0.08 &                Mohayaee and Tully (2005)\\
0.01 &                    0.70&                                   0.14&                  Ma and Scott (2013)\\
0.01 &                    0.76&                                   0.13&                  Nusser and Davis (2011)\\
0.01 &                    0.71&                                   0.10&                  Park and Park (2006)   \\
0.05 &                    0.71&                                   0.05&                  Kowalski et al. (2008)\\
0.06 &                    0.68&                                   0.10&                  Henry et al. (2009)     \\      
0.07 &                    0.75&                                   0.02&                  Beutler et al. (2012)\\
0.1  &                     0.71&                                   0.03&                  Tinker et al. (2012)\\
0.1  &                    0.71 &                                  0.03 &                 Simha and Cole (2013)\\
0.1  &                     0.72&                                  0.03 &                  Cacciato et al. (2013)\\
0.1  &                    0.76 &                                 0.02  &                  Cole et al. (2005)\\
0.1  &                     0.78&                                  0.02 &                  Sanchez et al. (2006)\\
0.16 &                    0.72 &                                 0.08  &                 Hicken et al. (2009)\\
0.2  &                     0.72&                                   0.07&                  Rozo et al. (2010)\\
0.2  &                     0.78&                                   0.08&                  Lampeitl et al. (2010)\\
0.2  &                     0.72&                                  0.02 &                  Hajian et al. (2013)\\
0.22 &                    0.72 &                                  0.05 &                 Blake et al. (2011a)\\
0.23 &                    0.73 &                                  0.03 &                Kessler et al. (2009)\\
0.275&                   0.71  &                                0.02   &               Percival et al. (2010)\\
0.3  &                     0.75&                                   0.05&                  Cabre and Cazta\~naga (2009)\\
0.3  &                     0.75&                                   0.07&                  Cao et al. (2012)\\
0.3  &                     0.80&                                   0.15&                  Rest et al. (2013)\\
0.3  &                    0.75 &                                  0.10 &                 Suyu et al. (2013)\\
0.3  &                    0.78 &                                  0.06 &                 Mantz et al. (2010)\\
0.3  &                    0.75 &                                   0.08&                 Vikhlinin et al. (2009)\\
0.3  &                    0.76 &                                   0.06&                Campbell et al. (2013)\\
0.3  &                    0.75 &                                   0.06&                Ho et al. (2008)\\
0.3  &                    0.73 &                                   0.02&                 Del Popolo et al. (2010)\\
0.35 &                   0.64  &                                  0.12 &                Zhang and Wu (2009)\\
0.35 &                   0.72  &                                  0.04 &               Reid et al. (2010)\\
0.35 &                   0.74  &                                  0.02 &               Sanchez et al. (2009)\\
0.35 &                   0.72  &                                  0.02 &               Mehta et al. (2012)\\
0.35 &                  0.76   &                                  0.04 &               Tegmark et al. (2006)\\
0.35 &                  0.73   &                                  0.04 &               Eisenstein et al. (2005)\\
 \end{tabular} 
\end{table}
\begin{table}[h] 
\caption{$\Omega_{\Lambda}$ table continued\label{lambdatable}} 
\begin{tabular}{l|r|r|l}
\hline
  \hline
 Redshift&       $\Omega_{\Lambda}$&       Uncertainty&        Reference  \\
   \hline  
0.36 &                   0.77  &                                  0.05 &               Okumura et al. (2008)\\
0.4  &                    0.85 &                                   0.08&                Taylor et al. (2012)\\
0.4  &                    0.72 &                                  0.06 &                Allen et al. (2008)\\
0.4  &                    0.73 &                                  0.04 &                Mandelbaum et al. (2013)\\
0.4  &                    0.82 &                                  0.03 &                Cabre et al. (2006)\\
0.4  &                   0.71  &                                  0.09 &               Blake et al. (2011b)\\
0.4  &                   0.76  &                                  0.20 &               Freedman et al. (2009)\\
0.41 &                  0.71   &                                  0.07 &               Blake et al. (2010)\\
0.43 &                  0.73   &                                  0.04 &               Wood-Vasey et al. (2007)\\
0.45 &                  0.85   &                                  0.15 &               Sullivan et al. (2011)\\
0.48 &                  0.73   &                                  0.15 &               Wilson et al. (2006)\\
0.5  &                   0.79  &                                   0.12&                Oguri et al. (2012)\\
0.5  &                   0.72  &                                   0.03&                Hinshaw et al. (2013)\\
0.5  &                   0.85  &                                  0.15 &               Cao and Zhu (2012)\\
0.5  &                   0.74  &                                  0.04 &               Ferramacho et al. (2009)\\
0.5  &                  0.71   &                                  0.06 &               Clocchiatti et al. (2006)\\
0.5  &                  0.71   &                                  0.02 &                Parkinson et al. (2012)\\
0.5  &                  0.72   &                                  0.02 &                Samushia et al. (2013)\\
0.52 &                0.75     &                                 0.10  &              Huff et al. (2011)\\
0.55 &                 0.75    &                                 0.10  &               Del Popolo (2010)\\
0.55 &                 0.70    &                                0.15   &              Ross et al. (2007)\\
0.55 &                 0.73    &                                0.04   &              Blake et al. (2009)\\
0.5  &                  0.70   &                                  0.10 &               Cazta\~naga (2005)\\
0.55 &                 0.73    &                                 0.04  &               Astier et al. (2006)\\
0.57 &                 0.74    &                                 0.05  &               Reid et al. (2012)\\
0.57 &                 0.74    &                                 0.04  &               Chuang et al. (2013)\\
0.57 &                0.72     &                                0.03   &               Zhao et al. (2013)\\
0.57 &                0.71     &                                0.02   &               Anderson et al. (2012)\\
0.57 &                0.72     &                                0.04   &               Chen et al. (2013)\\
0.59 &                0.70     &                                0.20   &               Farooq and Ratra (2013)\\
0.6  &                  0.69   &                                  0.13 &                 Blake et al. (2011a)\\
0.6  &                  0.79   &                                  0.10 &                 Guy et al. (2010)\\
0.6  &                  0.61   &                                  0.05 &                 Feoli et al. (2012)\\
0.6  &                 0.75    &                                  0.10 &                Conley et al. (2011)\\
0.6  &                  0.70   &                                  0.04 &                 Addison et al. (2013)\\
0.6  &                  0.64   &                                  0.12 &                Ganeshalingam et al. (2013)\\ 
0.6  &                  0.76   &                                 0.05  &                Lin et al. (2011)\\
 \end{tabular} 
\end{table}
\begin{table}[h] 
\caption{$\Omega_{\Lambda}$ table continued\label{lambdatable}} 
\begin{tabular}{l|r|r|l}
\hline
  \hline
 Redshift&       $\Omega_{\Lambda}$&       Uncertainty&        Reference  \\
   \hline   
0.6  &                  0.68   &                                0.05   &                Carneiro et al. (2006)\\
0.65 &                 0.73    &                                0.04   &               Davis et al. (2007)\\
0.66 &                 0.71    &                                0.02   &               Sereno and Paraficz (2014)\\
0.7  &                  0.84   &                                  0.10 &                 Heymans et al. (2013)\\
0.7  &                  0.72   &                                 0.02  &                 Seljak et al. (2005)\\
0.7  &                  0.75   &                                0.02   &                Freedman et al. (2012)\\
0.75 &                 0.75    &                                0.07   &               Benson et al. (2013)\\
0.75 &                 0.73    &                                0.02   &               March et al. (2011)\\
0.78 &                 0.78    &                                0.09   &                Blake et al. (2011a)\\
0.78 &                 0.76    &                               0.15    &               Benjamin et al. (2007)\\
0.80 &                 0.65    &                                0.10   &                Ettori et al. (2009)\\
0.80 &                 0.73    &                                0.10   &                Giannantonio et al. (2008)\\
0.88 &                 0.72    &                               0.08    &                Chen and Ratra (2011)\\
0.9  &                  0.68   &                                0.08   &                 Semboloni et al. (2006)\\
0.95 &                 0.75    &                               0.05    &                Fu et al. (2008)\\
1.0  &                  0.73   &                                 0.05  &                 Jassal et al. (2010)\\
1.0  &                  0.73   &                                0.03   &                Firmani et al. (2006)\\
1.0  &                  0.76   &                                0.06   &                Basilakos and Plionis (2010)\\
1.02 &                 0.73    &                               0.02    &               Suzuki et al. (2012)\\
1.3  &                  0.75   &                                 0.06  &                 Wei et al. (2013)   \\     
1.3  &                  0.76   &                                0.09   &                 Lopez-Corredoira (2013)\\
1.3  &                  0.73   &                                0.02   &                 Suzuki et al. (2012)\\
1.4  &                  0.65   &                                0.15   &                da Angela et al. (2005a)\\
1.5  &                  0.75   &                                0.08   &                 da Angela et al. (2008)\\
1.6  &                  0.63   &                                0.13   &                 Kodama et al. (2008)\\
1.6  &                  0.64   &                                0.10   &                 Tsutsui et al. (2009)\\
2.3  &                  0.75   &                                0.10   &                 Busca et al. (2013)\\
2.5  &                 0.73    &                               0.06    &                 Busti et al. (2012)\\
3    &                   0.65  &                                 0.20  &                  Da Angela et al. (2005b)\\
3    &                   0.75  &                                 0.03  &                  Dunkley et al. (2009)\\
3    &                   0.69  &                                 0.02  &                  The Planck Collaboration: Ade et al. (2013)\\
3    &                   0.72  &                                 0.03  &                  Bennett et al. (2013)\\
3    &                   0.76  &                                 0.02  &                  Spergel et al. (2007)\\
3    &                   0.74  &                                 0.03  &                  Reichardt et al. (2009)\\
3    &                   0.73  &                                 0.02  &                  Komatsu et al. (2009)\\
3    &                   0.73  &                                 0.02  &                  Komatsu et al. (2011)\\
3    &                   0.75  &                                 0.12  &                  Wang (2008)\\
\end{tabular} 
\end{table}
\begin{table}[h] 
\caption{$\Omega_{\Lambda}$ table continued\label{lambdatable}} 
\begin{tabular}{l|r|r|l}
\hline
  \hline
 Redshift&       $\Omega_{\Lambda}$&       Uncertainty&        Reference  \\
   \hline  
3    &                   0.75  &                                0.05   &                  Liang et al. (2008)\\
3    &                   0.73  &                                0.12   &                  Schaefer (2007)\\
3    &                   0.61  &                                0.14   &                  Sherwin et al. (2011)\\  
3    &                   0.65  &                                0.15   &                 Pietrobon et al. (2006)\\
3    &                   0.70  &                                0.02   &                 Spergel et al. (2013)\\
3    &                   0.65  &                                0.10   &                 Sievers et al. (2013)\\
3    &                  0.74   &                                0.02   &                 Larson et al. (2011)\\
      
 \end{tabular} 
\end{table}

\section{Strong limit: Spinning black holes and no-hair theorem}
There is plenty of evidence in support of the existence of black holes having masses in the range from a few 
$M_{\odot}$ to a few $10^{10}\, M_{\odot}$ (see A. Fabian elsewhere in this book).
However, to be sure that these objects are indeed the singularities predicted by General Relativity, we have to ascertain that at least in one case the spacetime around the suspected black hole satisfies the no-hair theorems.
The black hole no-hair theorems 
state that an electrically neutral rotating black hole in GR
is completely described by its mass $M$ and its angular momentum $S$
. 
This implies that 
the multipole moments, required to specify the external metric of a black hole, are fully expressible in terms of
$M$ and $S$. It is important to note that the no-hair theorems apply only in General Relativity, and thus they are a powerful discriminator between General Relativity and various alternative theories of gravitation that have been suggested (Will 2006, Yunes and Siemens 2013, Gair et al. 2013).

 A  practical test was suggested by Thorne and Hartle (1985) and Thorne et al. (1986). In this test the quadrupole moment $Q$ of the spinning body is measured. If the spin of the body is $S$ and its mass is $M$, we determine the value of $q$ in
\begin{eqnarray}
Q = -q \frac{S^2}{Mc^2}.
\end{eqnarray}
For true black holes $q=1$, and for neutron stars and other possible bosonic structures $q > 2$ (Wex and Kopeikin 1999, Will 2008). In terms of the Kerr parameter $\chi$ and 
the dimensionless quadrupole parameter $q_2$ the same equation reads
\begin{eqnarray}
q_2 = - q {\chi^2},
\end{eqnarray}
where $q_2 = {c^4\, Q}/{ G^2\,M^3} $ and $\chi = {c\, S}/{ G \,M^2}$.

An ideal test of the no-hair theorem is to have a test particle in orbit around a spinning black hole, and to follow its orbit. Fortunately, there exists such a system in nature.  The BL Lacertae object OJ287 is a binary black hole of very large mass ratio, and it gives well defined signals during its orbit. These signals can be used to extract detailed information on the nature of the orbit, and in particular, to find the value of the parameter $q$ in this system.
 
The optical light curve of this quasar displays periodicities of 11.8 and 55 years (Sillanp\"a\"a et al. 1988, Valtonen et al. 2006), as well as a $\sim 50$-day period of outbursts (Pihajoki et al. 2013). A model which explains these cycles, as well as a wealth of other information on OJ287 is discussed in the next section. 
\begin{figure}[t] 
\begin{center}
\includegraphics[width=\columnwidth] {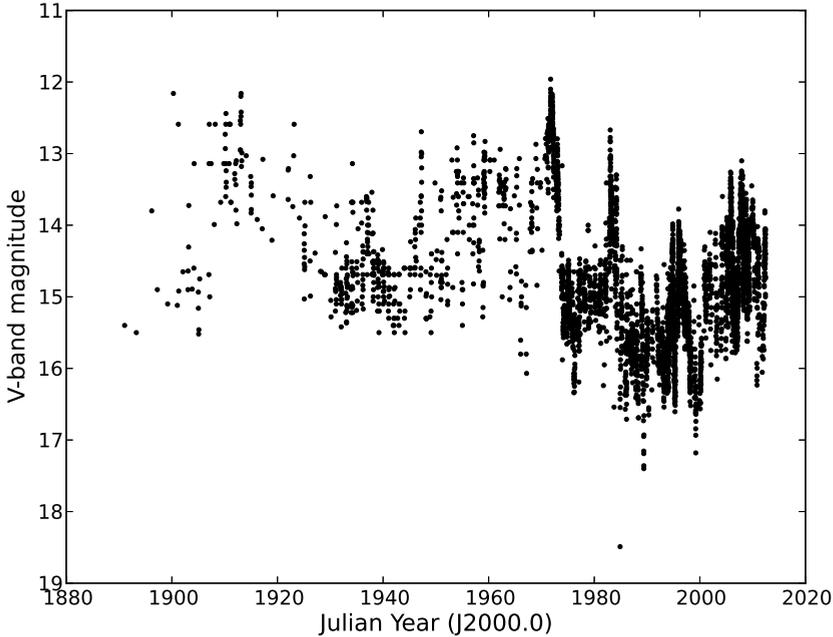}
\caption{The observation of the brightness of OJ287 from late 1800's until today.\label{fig1}} 
\end{center}
\end{figure} 

One of the pieces of information that we are able to find out is the spin of the primary black hole, 
$\chi_1=0.25 \pm 0.04$. The timing of the next outburst at the beginning of 2016 should 
help to improve the accuracy of $\chi$ to about $\pm 5 \%$. 
The mass of the primary is already determined with the 
accuracy of $\pm 1 \%$ which means that at least in principle we could reach the accuracy of $\pm 10 \%$ in measuring $q$ (Valtonen et al. 2011a).

 There exist a number of other proposals to test the black hole no-hair theorems. The scenarios include the radio timing of eccentric  millisecond binary pulsars which orbit an extreme Kerr black hole (Wex and Kopeikin 1999). Such systems are yet to be discovered. Also one may use several stars orbiting the massive Galactic center (Sgr A*) black hole at milliarcsec distances, if such stars are discovered and they are followed by infrared telescopes of the future, capable of doing astrometry at $\sim 10 \mu$ arcseconds level (Will 2008). Observations of gravitational waves from mergers of supermassive black holes, when they become possible some decades from now, may also be used to test the theorems (Barack and Cutler 2007).

 The imaging of accretion flow around Sgr A*, when it becomes possible, may allow the testing of
no-hair theorems. The test relies on the argument that a bright emission ring 
characterizing the flow image will be elliptical and asymmetric if the theorems are violated (Broderick et al. 2013). Finally, quasiperiodic oscillations, relativistically broadened iron lines, continuum spectrum and X-ray polarization in the accretion disk surrounding a spinning black hole may also be used as a probe of the no-hair theorems (Johannsen and Psaltis 2010a, 2010b, 2011, 2013, Bambi and Barausse 2011a, 2011b, Bambi 2012a, 2012b, 2012c, 2013, Krawczynski 2012).
  
It is expected that some of these tests should be possible by the middle of the next decade. Generally, there may be some difficulty due to the potential degeneracy between erraneous accretion physics close to the last stable orbit, which itself is not fully understood, and deviations from General Relativity (Broderick et al. 2013). In contrast, the test in OJ287 has already been done, and there are good prospects of improving the accuracy to $\pm10\%$ later in this decade. It does not depend on the accretion physics so close to the black hole. In what follows, we briefly summarize the current knowledge of the OJ287 system and the present constraints on the value of $q$.

\section{OJ287 binary system}
 The identification of the OJ287 system as a likely binary was made already in 1980's, but since the mean period of the system is as long as 12 yr, it has taken a quarter of century to find convincing proof that we are indeed dealing with a binary system (Valtonen 2008, Valtonen et al. 2008a, 2008b). The primary evidence for a binary system comes from the optical light curve. By good fortune, the quasar OJ287 was photographed accidentally since 1890's, well before its discovery in 1968 as an extragalactic object. The pre-1968 observations are generally referred to as ``historical'' light curve points. The light curve of over one hundred years (Figure 5) shows a pair of outbursts at $\sim12$ yr intervals. The two brightness peaks in a pair are separated by 1 - 2 yrs.

 The system is not strictly periodic, but there is a simple mathematical rule which gives all major outbursts of the optical light curve record. To define the rule, take a Keplerian orbit and demand that an outburst is produced at a constant phase angle and at the opposite phase angle. Due to the nature of Keplerian orbits, this rule cannot be written in a closed mathematical form, but the outburst times are easily calculated from it. According to this rule two outburst peaks arise per period. By choosing an optimal value of eccentricity (which turns out to be $e\sim0.7$) and by allowing the semimajor axis of the orbital ellipse to precess in forward direction at an optimal rate (which turns out to be $\Delta\phi\sim39^\circ\!$ per period), the whole historical and modern outburst record of OJ287 is well reproduced.

The type of model that follows from this rule is immediately obvious. It consists of a small black hole in orbit around a massive black hole. The secondary impacts the accretion disk of the primary twice during each full orbit (Figure 6).
\begin{figure}
 \includegraphics[width=\columnwidth]{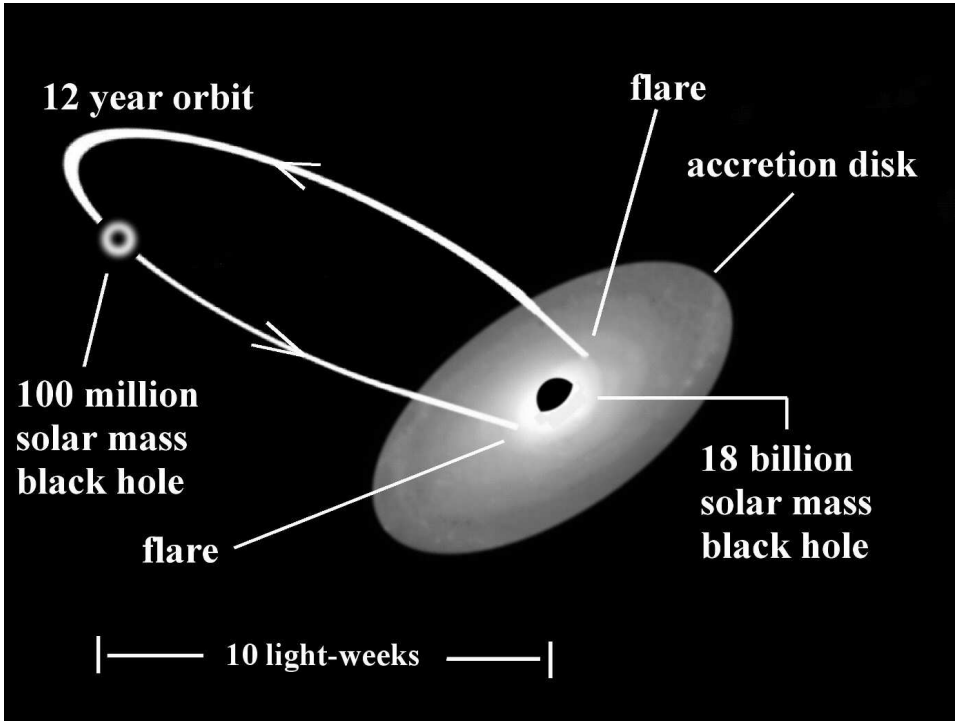} 
 \caption{An illustration of the OJ287 binary system. The jets are not shown, but they may be taken to lie along the rotation axis of the accretion disk. The two black holes are not resolved in current observations; the required resolution is $\sim10\mu$arcsec. However, the model explains \emph{all} observations from radio to X-rays and the time variability of these data.}
 \label{fig2}
\end{figure}
The two impacts produce the two flares that are observed 1 - 2 years apart, and the flares are repeated in every orbital cycle. However, because of relativistic precession, the pattern of flares is not exactly periodic. It is this fact that allows the determination of the precession rate, and from there the mass of the primary in a straightforward way. Note that it is not necessary to know the inclination of the secondary black hole orbit nor the orientation of the system relative to the observer in order to carry out these calculations.

The orbit that follows from the observed mathematical rule can be solved as soon as 5 flares are observed. Five flares have four intervals of time as input parameters, and the solution gives four parameters of the orbit uniquely: primary mass $m_1$, orbit eccentrity $e$, precession rate $\Delta\phi$, and the phase angle $\phi_0 $ at a given moment of time. There does not need to be any solution at all if the basic model is not correct. However, Lehto and Valtonen (1996) found a solution from five flares which already proved the case in the first approximation.

\begin{figure}
\includegraphics[width=4.5cm,angle=270]{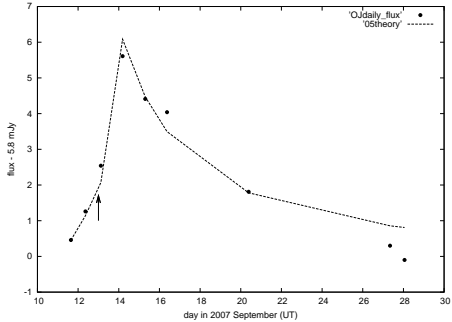} 
 \caption{The optical light curve of OJ287 during the 2007 September outburst. Only low polarization (less than $10\%$) data points are shown. The dashed line is the theoretical fit. The arrow points to September 13.0, the predicted time of origin of the rapid flux rise.}
\end{figure} 
At a deeper level one has to consider the astrophysical processes that generate the flares. The flares start very suddenly, with the rise time of only about one day (Figure 7). This fact alone excludes many possibilities that one might be otherwise tempted to consider: Doppler boosting variation from a turning jet or increased accretion rate due to varying tidal force. The timescales associated with these processes are months to years quite independent of a detailed model. It turns out that the base level of emission of OJ287, which is synchrotron radiation from the jet, is affected by both of these mechanisms. The Doppler boosting variation accounts for the 55 year cycle (Valtonen and Pihajoki 2013) while the varying tides change the base level in a 11.8 year cycle (Sundelius et al. 1997). Also the $\sim50$ day cyclic component is due to tides (Figure 8), but via a density wave at the innermost stable orbit of the primary accretion disk (Pihajoki et al. 2013).
\begin{figure}
\includegraphics[width=4.5cm,angle=270]{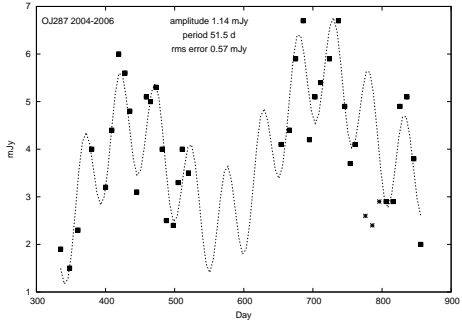} 
 \caption{The optical light curve of OJ287 during 2004-2006, using 10 day averages (squares). The 2005 impact flare is excluded. The three stars are points which are not used in the periodicity fit. The best fit for the remaing data is a 51.5 day period. If interpreted as the half-period of the innermost stable orbit, this implies $\chi_1=0.25$.}
\end{figure}
\begin{figure}
\includegraphics[width=4.5cm,angle=270]{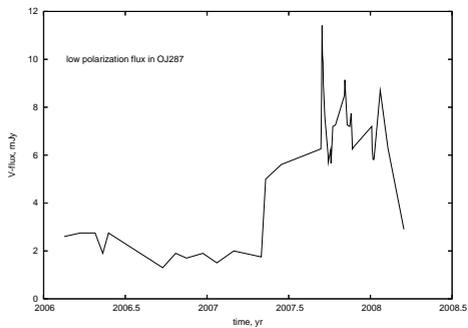} 
 \caption{The optical light curve of OJ287 during 2006-2008. Only low polarization (less than $10\%$) data points are shown.}
\end{figure}

 Figure 9 illustrates the light curve ``hump'' arising from increased accretion during the 2007 close approach of the two black holes. This feature is well modeled by simulations, and its time scale is about one year. In the same timescale the impact flare is a sharp ``spike'' on top the hump.

The correct astrophysical explanation of the flares is that the impact on the accretion disk releases hot bubbles of gas from the disk (Ivanov et al. 1998). These bubbles expand until they become optically thin, and the radiation from the whole volume is seen. The calculated light curve profiles agree with this model (Figure 7). The radiation is thermal bremsstrahlung at the temperature of about  $3\times10^5$ K (Valtonen et al. 2012). The addition of an unpolarised component to the emission lowers the degree of polarization (Valtonen et al. 2008b) which is another piece of evidence pointing to bremsstrahlung radiation. In contrast, the radiation from the ``hump'' is clearly synchrothron radiation with a raised number density of radiating particles (Seta et al. 2009). OJ287 is the only quasar which is known to have bremsstrahlung flares, and as such these flares give a unique set of signals to be used in orbit determination.
 
What is more important, the model is able to predict future outbursts. The prediction for the latest outburst was 2007 September 13 (Valtonen 2007, 2008a), accurate to one day, leaving little doubt about the capability of the model (see Figure 7; the predicted time refers in this case to the start of the rapid flux rise.) The 2007 September 13 outburst was an observational challenge, as the source was visible only for a short period of time in the morning sky just before the sunrise. Therefore a coordinated effort was made starting with observations in Japan, then moving to China, and finally to central and western Europe. The campaign was a success and finally proved the case for the binary model (Valtonen et al. 2008b).

 We see from Figure 7 that the observed flux rise coincides within 6 hours with the expected time. The accuracy is about the same with which we were able to predict the return of Halley's comet in 1986!

The astrophysical model introduces a new unknown parameter, the thickness of the accretion disk. For a given accretion rate which is determined from the brightness of the quasar (considering the likely Doppler boosting factor), the thickness is a function of viscosity in the standard  $\alpha$ disk theory of Shakura and Sunyaev (1973) and in its extension to magnetic disks by Sakimoto and Corotini (1981). The value of the viscosity coefficient $\alpha\sim0.3$ is rather typical for other accreting systems (King et al. 2007). Different values of $\alpha$ lead to different delay times between the disk impact and the optical flare. The delay can be calculated exactly except for a constant factor; this factor is an extra parameter in the model. The problem remains mathematically well defined. In fact, using only 6 outbursts as fixed points in the orbit, it is possible to solve the four orbital parameters plus the time delay parameter (in effect $\alpha$) in the first approximation (Valtonen 2007). The success of this model in predicting the 2007 outburst was encouraging.

The future optical light curve of OJ287 was predicted from 1996 to 2030 by Sundelius et al. (1997); they published the expected optical flux of OJ287 for every two-week interval between 1900 and 2030. During the first fifteen years OJ287 has followed the prediction with amazing
accuracy, producing five outbursts at expected times, of expected light curve
profile and size. It is extremely unlikely that such a coincidence should have happened by chance: these optical flux variations alone have excluded alternative models such as quasiperiodic oscillations in an accretion disk at the $5 \sigma$ confidence level (Valtonen et al. 2011b).

\begin{figure}
\begin{center}
\includegraphics [width=3in]{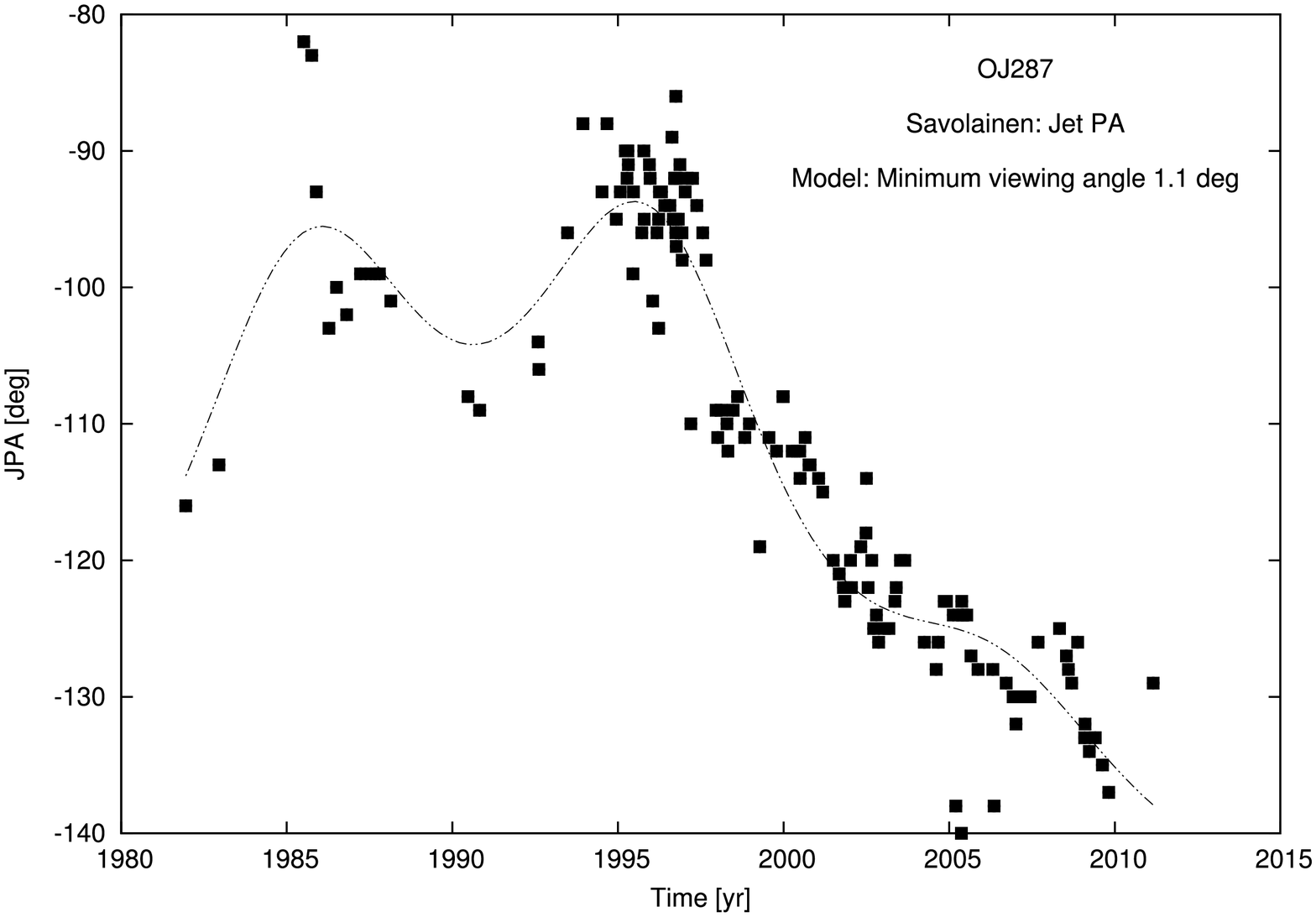} 

\end{center}
\caption{\label{label}Observations of the jet position angle in OJ287 at cm wavelengths. The line represents a model where the orientation change propagates outward in the jet with speed $0.85 c$. In observer's time, it takes over 200 years to propagate the change from the optical core to the cm-wave jet. Due to the small viewing angle of $\sim2^{\circ}$ of the jet, the small changes in its direction are magnified.}
\end{figure}
OJ287 is unresolved in optical, but in radio and X-ray wavelengths we observe a long jet. The radio jet has been observed since early 1980's and its orientation shows interesting, but in no way simple 12 year cycles. In Figure 10 we show the position angle (PA) observations of the jet as a function of time (Valtonen and Wiik 2012). The line follows a model where the binary shakes the accretion disk gravitationally. This shaking is transmitted to the jet. With suitably chosen parameters the binary action explains not only the cm-wave observations of Figure 10, but also quite different mm-wave jet wobble, in addition to the changes in the optical polarization angle (Valtonen and Pihajoki 2013). All these phenomena occur in the jet, far from the primary sites of action, and thus they cannot be used for high precision determination of the orbit.

The orbital parameters may be determined from the basic binary model without further knowledge of the orbit solution. The precession rate may be estimated by taking the ratio of the two dominant variability frequencies in the optical light curve, averaged over one month to minimize the effects of the impact flares on the analysis. If one of the frequencies (11.8 yr) relates to the orbital period while the other frequency (55 yr) arises from precession (Valtonen et al. 2006), then their ratio will tell what fraction of the angle $\pi$ the major axis of the orbit precesses per period. Due to the symmetry of the accretion disk relative to its midplane, the precessional effects should repeat themselves after rotation by $180^{\circ}$. Therefore the precession rate per orbital period should be $11.8\times180/55=38.\!^{\circ}6$. Once the precession rate is known, we get a good first estimate for the mass of the primary  $m_1\sim1.8\cdot 10^{10} M_\odot$.

The secondary mass was determined by Lehto and Valtonen (1996),  $m_2$ = $1.44\cdot 10^{8} M_\odot$, when adjusted to the 'current' Hubble constant of 70 km/s/Mpc and to the 5.6 mJy outburst strength, as observed in 2007. The mass value is based on the astrophysics of the impact and the strength of the maximum signal (Ivanov et al. 1998, Valtonen et al. 2012). A large mass ratio is necessary for the stability of the accretion disk. The primary black hole needs to be at least $\sim130$ times more massive than the secondary (Valtonen et al. 2012). The values of $m_1$ and $m_2$ satisfy this requirement.

The moderately high eccentricity $e\sim0.7$ is expected at a certain stage of inspiral of binary black holes of large mass ratio (Baumgardt et al. 2006, Matsubayashi et al. 2007, Iwasawa et al. 2011). Dynamical interaction with the stars of the galactic nucleus drives the eccentricity to $e\sim0.99$ before the gravitational radiation takes over in the evolution of the binary major axis. When the binary is at the evolutionary stage where OJ287 is now, the eccentricity has dropped to $e\sim0.7$.

The value of $\chi_1$ may be determined from the period of the innermost stable orbit. The best determination for this period is $103\pm4$ days which corresponds to  $\chi_1=0.25\pm0.04$.

The first two columns of Table 5 summarize the parameters determined by astrophysics of the binary black hole system. No information on the outburst timing has been used in the astrophysical model.

\begin{table}[h] 
\caption{Solution parameters.\label{solutiontable}} 
\begin{tabular}{l|r|r}
\hline
  \hline
   Parameter&Astrophysics&Orbit\\
   \hline 
$\Delta \phi$ & $38.\!^\circ6 \pm 1.\!^\circ0$ & $39.\!^\circ1 \pm 0.\!^\circ1$ \\ 
$m_1$&  $(1.8\pm 0.1)\cdot 10^{10} M_\odot$ & $(1.84\pm 0.01)\cdot 10^{10} M_\odot$\\ 
$m_2$&$(1.4\pm 0.4)\cdot 10^8 M_\odot$ & $(1.46\pm 0.1)\cdot 10^8 M_\odot$\\ 
$\chi_1$ & $0.25\pm 0.04$&  $0.28\pm 0.03$\\ 
$\phi_0$ & $56.\!^\circ0\ \pm 4.\!^\circ0$& $ 56.\!^\circ3\ \pm 1.\!^\circ0$\\ 
$e $ & $0.7\pm 0.03$& $0.70\pm 0.001$\\ 
$\alpha$& $0.3\pm 0.2$& $0.3\pm 0.1$\\ 
$q$ &--& $1.0\pm 0.3$\\ 

\end{tabular} 
\end{table}

\section{Modeling with Post Newtonian methods}

The Post Newtonian approximation to General Relativity provides the equations of motion
of a compact binary as corrections to the Newtonian equations of motion
in powers of $(v/c)^2 \sim G M / (c^2 r)$,
where $v$, $M$, and $r$ are
the characteristic orbital velocity,
the total mass, and the typical orbital separation of the binary,
respectively. The approximation may be extended to different orders.
The terminology 2PN, for example, refers to corrections to Newtonian dynamics in powers of $(v/c)^4 $. Valtonen et al. (2010a) use the 2.5PN and Valtonen et al. (2011a) 3PN-accurate orbital dynamics that includes the leading order 
general relativistic, classical spin-orbit and radiation reaction effects 
for describing the evolution of a binary black hole (Kidder 1995).
The following differential equations describe the relative acceleration of the binary 
and the precessional motion for the spin of the primary black hole at 2.5PN order: 

\begin{eqnarray} 
\ddot { {\vek x}} \equiv 
\frac{d^2  {\vek x}} { dt^2} &=& 
\ddot { {\vek x}}_{0} + \ddot { {\vek x}}_{1PN} \nonumber 
+ \ddot { {\vek x}}_{SO} +  \ddot { {\vek x }}_{Q}\\ 
&& + \ddot { {\vek x}}_{2PN} +  \ddot { {\vek x}}_{2.5PN}\,,  \\
\frac{d {\vek s}_1}{dt} &=&    {\vek \Omega}_{SO}  \times {\vek s}_1 \,,  
\end{eqnarray}  
\noindent
where ${\vek x} = {\vek x}_1 - {\vek x}_2 $ stands for the 
center-of-mass 
relative separation vector between the black holes with masses $m_1$ and $m_2$ and 
$  \ddot { {\vek x}}_{0}  $ represents the Newtonian acceleration given by
\begin{eqnarray}   
 \ddot { {\vek x}}_{0} = -\frac{ G\, m}{ r^3 } \, {\vek x} \,,
\end{eqnarray} 
 $m= m_1 + m_2$ and $ r = | {\vek x} |$. Kerr parameter $\chi_1$ and the unit vector 
${\vek s}_1$ define the spin of the primary black hole by the relation 
${\vek S}_1 = G\, m_1^2 \, \chi_1\, {\vek s}_1/c$ 
and $\chi_1$ is allowed to take values between $0$ and $1$ in general relativity. $\Omega_{SO}$ provides the spin precessional frequency due to spin-orbit coupling.  
The PN contributions occurring at the conservative 1PN, 2PN and the reactive 2.5PN orders, denoted 
by $\ddot { {\vek x}}_{1PN}$, $\ddot { {\vek x}}_{2PN}$ and 
$\ddot { {\vek x}}_{2.5PN}$, respectively, are non-spin by nature. The explicit expressions for these 
contributions, suitable for describing the 
binary black hole dynamics, are (Mora and Will 2004)
\begin{eqnarray}
\ddot { {\vek x}}_{1PN} =& 
- {\frac{ G\,m }{c^2\, r^2}} \biggl\{\hat {\vek n} \left[ -2(2+\eta) \frac{G\,m }{r} \right.  &\\ \nn  
& \left. + (1+3\eta)v^2 - \frac32 \eta \dot r^2  \right] 
  -2(2-\eta) \dot r {\vek v} \biggr\} \,, 
\end{eqnarray}
\begin{eqnarray}
\ddot { {\vek x}}_{2PN} 
=& - \frac{G\,m }{c^4\, r^2} \biggl\{\hat {\vek n} \biggl[ \frac34 (12+29\eta)\left(\frac{G\, m}{ r}\right)^2&\\ \nn 
&+ \eta(3-4\eta)v^4 + \frac{15}{8} \eta(1-3\eta)\dot r^4 & \\ 
& - \frac{3}{2} \eta(3-4\eta)v^2 \dot r^2 - \frac12 \eta(13-4\eta) (\frac{G\,m}{r})\, v^2&\\ \nn 
&- (2+25\eta+2\eta^2) ( \frac{G\,m}{r})\, \dot r^2 \biggr] & \\ \nn 
& - \frac12 \dot r {\vek v} \left[ \eta(15+4\eta)v^2 - (4+41\eta+8\eta^2) (\frac{G\,m}{r})\right. &\\\nn 
&\left. -3\eta(3+2\eta) \dot r^2 \right] \biggr\} \,, \label{a2PN} & 
\eea\bea
\end{eqnarray}

 
\begin{eqnarray}
\ddot { {\vek x}}_{2.5PN}  =& \frac{8}{15} \frac{ G^2 m^2 \eta }{ c^5 r^3 } 
\biggl \{ \left[ 9 v^2 + 17 \frac{G\,m}{r} \right] \dot{r} \hat{\vek n}& \\  \nn 
&- \left[ 3 v^2 + 9 \frac{G\,m}{r} \right] { \vek v} 
\biggr \} 
\,,& 
\end{eqnarray}
where the vectors $\hat {\vek n}$ and ${\vek v}$ are defined to be 
$ \hat {\vek n} \equiv {\vek x}/r $ and 
$ {\vek v} \equiv d {\vek x}/dt $, respectively, while 
$ \dot r \equiv dr/dt = \hat {\vek n} \cdot {\vek v}$, 
$ v \equiv | {\vek v} |$ and the symmetric mass ratio $\eta = m_1\, m_2/m^2$.

The leading order spin-orbit contributions to $ \ddot {{\vek x} }$, 
appearing at 1.5PN order (Barker and O'Connell 1975a), read 
\begin{eqnarray} 
\ddot { {\vek x}}_{SO} &= \frac{ G\, m}{ r^2} \, \left ( \frac{G\,m}{c^3\, r} \right )\, 
\left ( \frac{ 1 + \sqrt{1 -4\,\eta} }{4} \right )&\\ & \nn 
\, \chi_1 
\biggl \{ \biggl [ 12 \, \left [ {\vek s}_1 \cdot ( \hat {\vek n} \times {\vek v} ) \right ] \biggr ]\, 
\hat {\vek n} \\ & 
+ \biggl [ \left ( 9 + 3\, \sqrt {1 - 4\, \eta} \right ) \, \dot r \biggr ] 
\left ( \hat {\vek n} \times {\vek s}_1 \right ) 
\nn \\ & 
- \biggl [ 
7 +  \sqrt {1 - 4\, \eta} 
\biggr ] 
\left ( {\vek v} \times {\vek s}_1 \right ) 
\biggr \}\,,\nn 
\end{eqnarray}
while
\begin{eqnarray} 
{\vek \Omega}_{SO} = \left ( \frac{G\,m\,\eta}{2c^2\, r^2} \right )  
\biggl ( \frac{ 7 + \sqrt {1 - 4\, \eta} }{ 1 + \sqrt {1 - 4\, \eta} } 
\biggr ) 
\left ( \hat {\vek n} \times \hat  {\vek v} \right )\,.\\ 
\end{eqnarray}

Finally, the quadrupole-monopole interaction term $\ddot { {\vek x}}_Q $, entering at the 2PN order (Barker and O'Connell 1975a), reads
\begin{eqnarray} 
\ddot { {\vek x}}_Q & =- q \, \chi_1^2\, 
\frac{3\, G^3\, m_1^2 m}{2\, c^4\, r^4}
\, \biggl \{ 
\biggl [ 5(\hat{\vek n}\cdot {\vek s_1})^2 
-1 \biggr ]\hat {\vek n}
\nn \\&
-2(\hat{\vek n}\cdot \vek s_1) {\vek s_1} \biggr \}, 
\end{eqnarray} 
where the parameter $q$, whose value is $1$ in general relativity, is introduced to test the black hole `no-hair' 
theorems (Will 2008). 

   It turns out that adding the 3PN contributions to $d^2{ {\vek x}}/dt^2 $, are not necessary; their influence falls within error limits of the OJ287 problem. It has been important to verify that the PN series converges rapidly enough to ingore the terms of order higher than 2.5PN. Similarly, the terms related to the spin of the secondary turn out be negligle in the OJ287 problem (Valtonen et al. 2011a). The precessional motion for the spin of the primary black hole is 2PN accurate in our calculations.


  \begin{figure}[ht] 
\includegraphics[angle=0, width=3in] {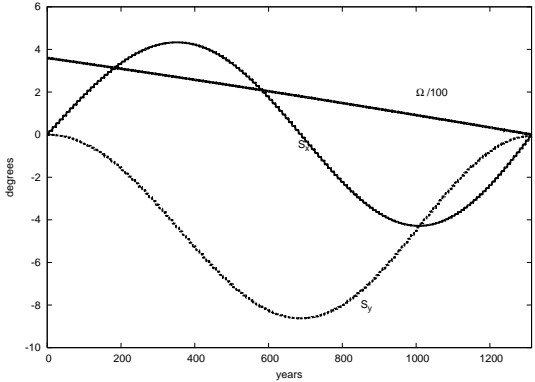}  
  \caption{
  The circulation of the ascending node ( $\Omega$) of the orbit 
   (straight line) in the coordinate system in which the initial 
   spin vector $\vek s_1$  points to the direction of the z-axis. 
   The quantities plotted are  $\Omega/100$ in degrees and 
   the symbols $s_x$ and $s_y$ correspond to the $x$ and $y$ 
   components of the spin-direction vector $\vek s_1$ in degrees.\label{fig4}. 
} 
\end{figure}

  The main consequence of including the leading order spin-orbit interactions to the dynamics 
of a binary black hole is that it forces the orbital plane to precess. The orbital angular momentum vector, characterising the orbital plane, precesses around 
the spin of the primary in such a way that the angle between the orbital plane and 
the spin vector ${\vek s}_1 $ remains almost constant (roughly within $\pm 0.\!^\circ5$ in the model). 
The spin-vector itself precesses drawing a cone with an opening angle of about 8 degrees 
(see Figure 11).

The precessional period for both the orbital plane and the spin of the binary, provided by $|{\vek \Omega}_{SO}|$, 
is about 1300 years for the orbital parameters of OJ287. The precession cone axis coincides with the mean accretion disk axis. 
 It is 
reasonable to expect such a situation due to the Bardeen-Petterson effect (Kumar and Pringle 1985). Because the time scale of the Bardeen-Petterson effect is much longer than the black hole spin 
precession time scale (Lodato and Pringle 2006), the disk axis is not able to follow the rapid precession of the primary spin axis.

The orbit solutions currently involve  {\it nine} 
accurately timed outbursts as listed in Table 6.
The orbit solutions provide a unique set of parameters (Table 5, column 3).
 Recently also the tenth outburst from 1906 has been added (Hudec et al. 2013). In this case there are only limits for the timing; the main influence of including the 1906 data is to bring $\chi_1$ to $\sim0.23$, somewhat below the range given in Table 5. Note that the eccenticity $e$ is defined as
\begin{eqnarray}
\frac{r_a}{r_p} = \frac{1+e}{1-e},
\end{eqnarray}
where $ r_a$ and $r_p$ are the apocenter and pericenter distances in the orbit, respectively. The instantaneous eccentricity varies during a relativistic binary orbit; in this way we get a definite number that resembles the eccentricity of a Keplerian orbit.


\begin{table}[t] 
\caption{Outburst times with  estimated uncertainties\label{outburst}. 
These are starting times of the outbursts.} 
\begin{tabular}{lr}
\hline 
  \hline
   Time&uncertainty\\
   \hline 
1912.970& $\pm$  0.010\\ 
1947.282& $\pm$  0.0005\\ 
1957.080& $\pm$  0.020\\ 
1972.94& $\pm$  0.005\\
1982.964&$\pm$   0.0005\\  
1984.130&$\pm$   0.002\\ 
1995.843&$\pm$   0.0005\\ 
2005.745& $\pm$  0.005\\ 
2007.692&$\pm$   0.0005\\ 
\end{tabular} 
\end{table}

Because we are using 8 time intervals to fit 8 parameters, the fit has no degrees of freedom. The tolerance limits of Table 5 are a consequence of having a tolerance in the fixed points of the outburst times (Table 6). The no-hair parameter $q$ is the eigth parameter which of course should be $q=1$ if General Relativity is correct.

As part of the solution we obtain the list of all past and future outburst times with their uncertainties. For example, the well recognized outbursts in 1959, 1971 and 1994 are timed at $1959.213\pm0.002$, $1971.1265\pm0.002$
and $1994.6085\pm0.005$, 
respectively. In all these cases data are missing at the crucial time of expected rapid flux rise, 
and thus these predictions cannot be verified at present.

 It is remarkable that the two sets of parameters in second and third columns of Table 1 agree so closely. If the basic model were not correct, there would be no reason for this agreement. Also, the mere fact that an orbit solution exists at all is a strong argument for the model.

  Let us now turn our attention to the expected future outbursts in OJ287.
We expect three more outbursts during the next two decades, occuring in 2016, 2019 and 2022. 
The 2016 outburst should be an easy one to detect, as it comes in January of that year. 
Its timing is spin-sensitive; the exact date will 
give us a good spin value from the following formula:
\begin{eqnarray}
\chi_1 = 0.25-0.5\times(t-2016.0).
\end{eqnarray} 

Here $t$ is the time of the beginning of the outburst in years. The expected accuracy is 0.005 units in $\chi_1$. The dependence on $q$ is secondary, and thus the 2016 outburst timing is of no use by itself for the testing of the no-hair theorem.

The value of $q$ is currently best determined by the 1995 outburst. There was an intensive monitoring campaign of OJ287 (called OJ94) during this outburst season, 
but unfortunately there exists a gap in these observations just at the crucial time (Valtonen 1996, Figure 12). 
It may still be possible that there are measurements somewhere which are not recorded in the OJ94 campaign light curve, 
and which would be valuable in narrowing down $q$ even from these data. The line in Figure 12 is drawn using the well 
observed 2007 outburst as a template to compare with 1995. It is should be noted that 
even a few more measurements of 1995 could narrow down the range of $q$.


\begin{figure}[ht] 
\includegraphics[angle=270, width=3in] {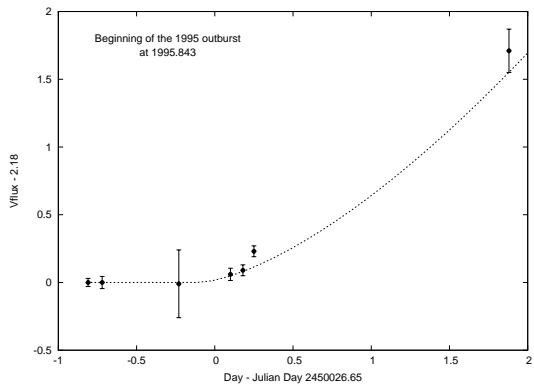}
\caption{ 
 The observations of OJ287 during 4 - 7 November 1995, transformed to optical V-band. Overlaid is the theoretical light curve profile, corresponding to  $q=1$. The zero point of the
time axis is at 3:36 UT of 1995, November 5. The peak flux of 5.1 mJy above the base level was predicted for November 8 (Lehto and Valtonen 1996), but there were no observations to confirm it. The data on November 9 and later suggest that the outburst maximum was probably missed.\label{fig3} } 
\end{figure}

  The 2019 outburst timing is also sensitive to the $q$:
\begin{eqnarray}
q = 1.0 - 1.2\times(t-t_0),
\end{eqnarray}
where $t$ is time of the beginning of the outburst (in days) and $t_0$ is 2019 July 14, 11:19 UT. 
The expected accuracy for the $q$ determination is $0.15$ units, i.e. optimally the accuracy of the no-hair theorem test is doubled in comparison with the present uncertainty.
Observing OJ287 during the rising flux of the 2019 outburst, from July 14 to July 17, is 
a challenge for ground based observers.
This is because the angular distance between the Sun and OJ287 
is only about 19 degrees at the beginning of the event, and goes down to about 16 degrees by the time of the peak flux on July 17. T. Pursimo has measured OJ287 at the Nordic Optical telescope on 2007 July 12 (Valtonen et al. 2009) which demonstrates that the July measurements should also be feasible in 2019.
Space observations would also be useful in order to study the outburst over a wider spectral range.
The 2022 outburst is scheduled practically at the same time of year as the 2019 outburst. 
Obviously it would also be of interest to observe this event as it would tie down the parameters of the general model more narrowly. 
However, it will not give any further information on $q$.

\section{OJ287 results at the strong field limit} 
  The binary black hole in OJ287 is modeled to contain a spinning primary black hole with an accretion disk
  and a spinning or non-spinning secondary black hole. Using PN-accurate dynamics, relevant for such a system, 
we infer that the primary black hole 
should spin approximately at one quarter of the maximum spin rate allowed in general relativity. 
In addition, the `no-hair theorems' of black holes
(Misner et al. 1973)
are supported by the model, although the testing
is possible only to a limited precision. 
The solutions concentrate around $q=1.0$ with one standard deviation of $0.3$ units.
These results are achieved with the help of new data on historical outbursts as well 
as using the most recent outburst light curves together with 
the timing model for OJ287 outbursts (Valtonen et al. 2007). 

Further, a polar orbit is assumed. Any other high inclination model would give us the same timing results.
Sundelius et al. (1997) carried out simulations with a number of different inclinations between the disk and the orbit, and found that the inclination makes no difference. This is not surprising since the impacts occur along the line of nodes, and the timing of the impacts is not a function of the impact angle. We have tried varying the spin angle relative to the disk, and found that within reasonable limits (we considered inclinations up to $10^\circ$) the results do not change. Again this is what is expected since the line of nodes circulates slowly in comparison with the orbital angular motion.

The reason why we would expect at least approximate alignment between the black hole spin and the disk spin is the Bardeen - Petterson effect which tends to align the two spins. However, the alignment time scale is much slower (about $10^{7}$ yr, Lodato \& Pringle 2006) than the spin precession time, and thus the disk does not follow the black hole spin exactly but only on average.

In this scenario, we have a unique solution and also a unique prediction for the next OJ287 outburst, 
expected in 2016.  
We should then be able to judge the correctness of the present solution. Note
that an outburst is not expected in 2016 in any simple extrapolation from past observations, as it is well before the
average 12 yr cycle is due, and thus it is a sensitive test of the general model as well as a test 
for the spin of the primary black hole. The observations of the 2019 outburst will then improve the accuracy of the 'no-hair' theorem test by a factor of two, to the level of $15\%$.

There are at least 5 additional outbursts in the historical record which have not yet
been been accurately observed. If new data of these outbursts are found it will open up the possibility
of improving the model and improving that accuracy of the no-hair theorem test to the level of $10\%$. Searches of plate archives are recommended for
this purpose.
\section{Summary and Conclusions on General Relativity at weak and strong  field limits}

We have discussed the amazing range of validity  of General Relativity from tests from the weak to  strong field limits. The weak field corresponds to speeds much less than that of light or distances much larger than the Schwarzschild radius of the primary body of an orbiting pair.   In weak field tests, we briefly reviewed the  precession of Mercury's orbit, bending of light near the Sun, precession of a binary pulsar orbit, gravitational radiation from the binary pulsar, relativistic geodetic precession and the precession due to the relativistic Lense-Thirring effect. In all these cases General Relativity accounts for the observations but, the cosmological constant (manifested as dark energy) is not significant.  

However, the existence of dark energy in General Relativity can be tested in the weak field limit.  Einstein (1917) proposed a cosmological constant specified to produce a static universe. Evolving models of the universe were derived by Alexander Friedmann (1922, 1924) and Georges Lemaitre (1927).  The Einstein static solution is a special case, where the cosmological constant is related to the matter density to give gravity-antigravity balance.  In general there are no theoretical restrictions on the cosmological constant. The universe expands or contracts as a whole. The Hubble discovery of the redshift distance relation (Hubble 1929) can be regarded as a predictive verification of General Relativity.

After the discovery of the Big Bang, it seemed that  the cosmological constant  could not be as big as Einstein required for a static universe.  It was subsequently commonly assumed to be zero. However, when  supernovae of type Ia were used to estimate distances of galaxies whose redshifts $z\sim1$, a positive cosmological constant was indicated (Riess et al. 1998 and Perlmutter et al. 1999).

We describe in detail how to test the existence of dark energy at the weak field limit of General Relativity via an outflow model in groups and clusters of galaxies (Chernin 2001; Chernin et al. 2006; Byrd et al. 2011). The Local Group has been found to have outflowing dwarf galaxies around it (van den Bergh 1999; Karachentsev et al. 2009). The dwarfs' motion indicates not only the gravitational mass of the group but also the dark energy background (Chernin et al. 2009). Studies the Virgo and Fornax clusters and even rich clusters like Coma (Karachentsev et al. 2003, Chernin 2008, Chernin et al. 2006, 2007, 2010, 2012a,b, Chernin 2013) show similar structure.

  There has been been a large amount of effort to determine the value of the cosmological constant, and its possible dependence on redshift or, equivalently, time (Frieman et al. 2008, Blanchard 2010, Weinberg et al. 2013). In the literature from year 2005 onwards, values cluster around $\Omega_{\Lambda}  = 0.73$ with a standard deviation of 0.044, the typical error in individual measurements. Only the cosmic microwave background (CMB) models give a significantly better accuracy, but even in this case one may suspect hidden systematic errors due to foreground corrections (Whitbourn et al. 2014). In our analysis  of the literature including nearby groups and clusters, within the error limits,  $\Lambda$   appears to be a constant over the redshift range from cosmological  to Local Group outflows, an impressive validation.

The strong field corresponds to speeds comparable to that of light or distances comparable to the Schwarzschild radius of the primary body of a pair.  An important strong field test  is to be sure that black holes are indeed the singularities predicted in General Relativity. We have to ascertain that at least in one case the spacetime around a suspected black hole satisfies the no-hair theorems which state that an electrically neutral rotating black hole in General Relativity is completely described by its mass, $M$, and its angular momentum, $S$,  implying that the multipole moments, required to specify the external metric are fully expressible in terms of $M$ and $S$. No-hair theorems apply only in General Relativity, and thus are a powerful eliminator of  various alternative theories of gravitation  (Will 2006, Yunes and Siemens 2013, Gair et al. 2013).

An ideal test of the no-hair theorem is to have a test particle in orbit around a spinning black hole.  We have described results for one such system, the BL Lacertae object OJ287, a binary black hole of very large mass ratio. OJ287 gives outbursts during the secondary's periodic orbital disk crossings of an accretion disk around the more massive member. The timing of the repeating outbursts gives the orbit and the primary's quadrupole parameter. Post-Newtonian dynamics, necessary for such a system, implies that the primary black hole should spin at about one quarter of the maximum spin rate allowed in General Relativity.  The 'no-hair theorems' of black holes (Misner et al. 1973) are supported by the model but with a limited precision at present (Valtonen et al. 2011a).  Observations of predicted 2016 and 2019 outbursts and collection of data on more past outbursts should improve the accuracy of the 'no-hair' theorem test. This extreme test at the strong field limit supports General Relativity.

\end{document}